\documentclass[10pt,preprint]{aastex}

\slugcomment{\today}

\newcommand{\eg}{\mbox{e.g.}}
\newcommand{\ie}{\mbox{i.e.}}

\newcommand{\minusone}{$^{-1}$}

\newcommand{\nh}{\mbox{$N_{\rm H}$}}

\newcommand{\skipthis}[1]{}

\def\micron{\hbox{$\mu$m}}
\newcommand{\be}{\begin{equation}}
\newcommand{\ee}{\end{equation}}
\newcommand{\e}{et al.\ }

\def\ie{{i.e.\ }}

\usepackage{epsf}
\usepackage{textcomp}
\usepackage{url}
\usepackage{natbib}
\shorttitle{YSOVAR: Mid-IR Variability in GGD~12-15}
\shortauthors{Wolk \e}

\begin{document}

\title{YSOVAR: Mid-infrared Variability Among YSOs in the 
Star Formation Region GGD~12-15}

\author{Scott J. Wolk}
\affil{Harvard--Smithsonian Center for Astrophysics, 60 Garden
Street, Cambridge, MA 02138}
\email{swolk@cfa.harvard.edu}
\and
\author{H. Moritz G\"{u}nther}
\affil{Harvard--Smithsonian Center for Astrophysics, 60 Garden
Street, Cambridge, MA 02138}
\affil{MIT Kavli Institute, 70 Vassar Street, Cambridge, MA 02139}
\and
\author{Katja Poppenhaeger}
\affil{Harvard--Smithsonian Center for Astrophysics, 60 Garden
Street, Cambridge, MA 02138}
\affil{NASA Sagan fellow}
\and
\author{A. M. Cody}
\affil{NASA Ames Research Center, M/S 244-5 Moffett Field, CA 94035, USA}
\affil{Spitzer Science Center, California Institute of Technology, Pasadena, CA 91125, USA}
\and
\author{L. M. Rebull}
\affil{Spitzer Science Center/Caltech, 1200 E. California Blvd., Pasadena, CA 91125, USA}
\and
\author{J. Forbrich}
\affil{University of Vienna, Department of Astrophysics, T\"urkenschanzstr. 17, 1180 Vienna, Austria}
\affil{Harvard--Smithsonian Center for Astrophysics, 60 Garden Street, Cambridge, MA 02138}
\and
\author{R. A. Gutermuth}
\affil{Dept. of Astronomy, University of Massachusetts, Amherst, MA  01003, USA}
\and
\author{L. A. Hillenbrand}
\affil{Department of Astronomy, California Institute of Technology, Pasadena, CA 91125, USA}
\and
\author{P. Plavchan}
\affil{Department of Physics Astronomy and Materials Science, Missouri State University,
Springfield, MO 65897}
\and
\author{J. R. Stauffer}
\affil{Spitzer Science Center/Caltech, 1200 E. California Blvd., Pasadena, CA 91125, USA}
\and
\author{K. R. Covey}
\affil{Dept. of Physics \& Astronomy, Western Washington Univ.
Bellingham, WA 98225-9164}
\and
\author{Inseok Song}
\affil{Physics and Astronomy Department, University of Georgia, Athens, GA 30602-2451, USA}

\skipthis{
\author{L. Allen}
\affil{National Optical Astronomy Observatories, Tucson, AZ, USA}
\and
\author{A. Bayo}
\affil{Max Planck Institut f\"ur Astronomie, K\"onigstuhl 17, 69117, Heidelberg, Germany}
\affil{Departamento de F\'isica y Astronom\'ia, Facultad de Ciencias, Universidad de Valpara\'iso, Av. Gran Breta\~na 1111, 5030 Casilla, Valpara\'iso, Chile }
\and
\author{J. L. Hora}
\affil{Harvard-Smithsonian Center for Astrophysics, 60 Garden Street, Cambridge, MA 02138, USA}
\and
\author{H. Y. A. Meng}
\affil{Infrared Processing and Analysis Center, California Institute of Technology, MC 100-22, 770 S Wilson Ave, Pasadena, CA 91125, USA}
\affil{Lunar and Planetary Laboratory, University of Arizona, 1629 E University Blvd, Tucson, AZ 85721, USA }
\and
\author{M. Morales-Calder\'on}
\affil{Centro de Astrobiolog\'ia (INTA-CSIC), ESAC Campus, P.O. Box 78, E-28691 Villanueva de la Canada, Spain}
\and
\author{J. R. Parks}
\affil{Department of Physics and Astronomy, Georgia State University, 25 Park Place South, Atlanta, GA 30303, USA}
}

\begin{abstract}
 We present an IR-monitoring survey with the $Spitzer$ Space Telescope of the star forming region GGD~12-15.  Over 1000 objects were monitored including about 350 objects within the central 5\arcmin\ which is found to be especially dense in cluster members.  The monitoring took place over 38  days and is part of the Young Stellar Object VARiability (YSOVAR) project.    The region was also the subject of a contemporaneous 67~ks $Chandra$ observation. The field includes 119 previously identified pre-main sequence star candidates.   X-rays are detected from 164 objects, 90 of which are identified with cluster members.  Overall, we find that about half the objects in the central 5\arcmin\ are young stellar objects based on a combination of their spectral energy distribution, IR variability and  X-ray emission.
 Most of the stars with IR excess relative to a photosphere show large amplitude ($>$0.1 mag)  mid-IR variability. There are 39 periodic sources, all but one of these is found to be a cluster member. Almost half of the periodic sources do not show IR excesses.  Overall, more than 85\% of the Class I, flat spectrum, and Class II sources are found to vary. The amplitude of the variability is larger in more embedded young stellar objects.   Most of the Class~I/II objects exhibit redder colors in a fainter state, compatible with time-variable extinction. A few become bluer when fainter, which can be explained with significant changes in the structure of the inner disk. A search for changes in the IR due to X-ray events is carried out, but the low number of flares prevented an analysis of the direct impact of X-ray flares on the IR lightcurves. However, we find that X-ray detected Class II sources have longer timescales for change in the mid-IR than a similar set of non-X-ray detected Class IIs. 
\end{abstract}


\section{Introduction}
\label{sec:Intro}
Circumstellar disks are an ubiquitous product of star formation, as revealed by the plethora of young stellar objects (YSOs) with mid-IR excesses detected by the \textit{Spitzer} Space Telescope \citep{All04, Chu09, Eva09}. These disks mediate mass accretion and angular momentum loss in pre-main sequence (PMS) stars, and also represent the immediate environment for planet formation. Understanding the physical properties and evolutionary paths of these disks is central to constructing viable models of the formation of stars and planets.
  
\subsection{Variability of Young Stars}
Photometric variability was one of the original defining characteristics 
of YSOs \citep{Joy46}. Optical monitoring is primarily sensitive to events 
and features near the stellar photosphere; such studies have determined 
the rotational periods of YSOs \citep[e.g.,][]{Ryd83} and the sizes, 
temperatures and temporal stability of hot and cold spots on YSO 
photospheres \citep[e.g.,][]{Bou86}.  Studies of individual YSOs such as 
AA Tau and its analogs reveal insights into magnetospheric accretion 
processes linked to inner-disk dynamics \citep{Bou03, Bou07, Don10}.  
Most variability studies have used optical monitoring.    However, the 
embedded nature of many YSOs favor near- and mid-IR wavelengths.
Near-infrared (NIR) studies of young stars allow for the direct 
detection of the hottest portions of optically thick protoplanetary disks 
around these stars via excess $K$-band flux \citep{Lad92}. 
Studies of the Taurus, Orion and Chameleon I molecular clouds
established that NIR variability is present in YSOs \citep{Skr96, Car01, Car02}. 
In Orion, \citet{Car01} identified more than a thousand YSOs with
NIR variability in their survey of the Orion Molecular cloud, and
established a strong connection 
between variability and near-infrared excess.

Longer baseline JHK studies on sources in Cyg OB7 and the Orion Nebular 
Cluster (ONC) find over 90\%  of YSOs vary significantly \citep{Ric12, Ric15}. 
Not only do YSOs have NIR colors that vary with time, but \citet{Ric12} 
also found the color of some stars changes enough to move 
them from the region of color-magnitude space attributed to stars with disks to the 
  region of color-magnitude space attributed to stellar photospheres alone.\footnote {The 
presence or absence of a disk is often inferred from the NIR colors 
following \citet{Lad92}.}  While some of the variations may be rotationally 
modulated surface spots (hot or cold), other color shifts were identified 
with changes in the disk structure including altering the accretion rate 
which is temporally modulated.  In a follow-up Cyg~OB7 paper, \citet{Wol13} 
divided the variability into classes based on their taxonomy and associated 
the classes with different dominant physical processes.   They found that 
in addition to reddening due to changes in the dust along the line of sight, 
some sources became bluer when fainter.  \citet{Wol13} suggested this 
blueing was due to changes in the accretion disk structure. Similarly, 
L1688 was monitored over a 2.5 year period in near-IR wavelengths with 
stars found to vary due to a combination of cool spots, variable accretion and 
disk eclipses \citep{Par14}. Also in the near-IR, the long term variability 
of stars in several young clusters including the ONC, NGC 1333, IC 348 
and $\sigma$ Orionis was studied by \citet{Sch12} who found that amplitudes 
are largest in NGC 1333, presumably because it has the youngest sample of 
YSOs. He also finds that the frequency of detecting objects as being
highly variable objects also increases with the time window of the observations.

While the JHK lightcurves measure the changes to the inner edge of the 
disk, mid-IR lightcurves are sensitive to changes in what is expected to 
be a more dynamically stable portion of circumstellar disk, slightly away from the 
inner edge. Because of its Earth-trailing orbit and very stable and sensitive 
detector arrays, the $Spitzer$ Space Telescope \citep{Wer04} has provided 
the first platform capable of useful synoptic monitoring of YSOs at mid-IR 
wavelengths.  Early results from \textit{Spitzer} monitoring demonstrated 
significant variability in mid-IR emission from many PMS disks. \citet{Mor09} 
carried out the first comprehensive \textit{Spitzer} survey of YSO mid-IR 
variability using the InfraRed Array Camera \citep[IRAC;][3.6 -- 8.0\micron]{Faz04} 
observing $\sim$ 65 members of the embedded, $\sim$1 Myr old, IC 1396A 
star-forming core twice a day for 14 days. Almost half of the YSOs were 
detectably variable; about half of those showed 4-12 day period-like behavior while 
the other half exhibited transient, non-periodic variability.

The Young Stellar Object Variability project \citep[YSOVAR;][hereafter 
Paper~I] {Reb14} has obtained extensive, multiwavelength time series 
photometry over 40-700 day intervals for $>$1000 YSOs in Orion and
in a number of smaller star-forming cores. YSOVAR 
observations of the ONC \citep{Mor11} revealed a number of dust-eclipse 
events similar to those seen in AA Tau \citep{Bou03}, giving insight into 
the structure and behavior of protoplanetary disks around stars of this 
age, as well as the importance of magnetically driven accretion onto young stars.  
\citet{Gun14} present YSOVAR data for 
Lynds 1688 covering several epochs.   They find almost all cluster members 
show significant variability and that  the amplitude of the variability 
is larger in more embedded YSOs. Like \citet{Wol13}, they find a significant 
fraction of sources become bluer as they become fainter -- except in this 
case the colors in question are mid-IR and not near-IR.  In a followup
program to YSOVAR, \citet{Cod14}, 
\citet{Sta14} and \citet{Sta15} present comprehensive analyses of simultaneous optical
lightcurves from the Convection, Rotation \& Planetary Transits (CoRoT) mission 
and infrared lightcurves from \textit{Spitzer} for sources in NGC~2264.  They 
focus on 162 classical T Tauri stars using metrics of periodicity, 
stochasticity, and symmetry to statistically separate the lightcurves 
into seven distinct classes, which they suggest represent a combination 
of different physical processes and geometric effects.  \citet{McG15}  
estimate that 14\% of the classical T Tauri stars observed in NGC~2264
are AA-Tau-like systems, where
the optical variability is primarily due to periodic occultation of our
line of sight to the YSO by a warped inner disk.  Based on additional 
spectroscopic and photometric data, the AA Tau analogs are believed to
generally be undergoing relatively stable, ``funnel-flow" accretion along
the stellar dipole magnetic field lines \citep{Rom11}.  The optical 
light curves of the most heavily accreting stars in NGC~2264 were generally
found by \citet{Sta15} to be dominated by short-duration flux bursts, most
plausibly associated with the 
unstable accretion mode seen in some 3D MHD simulations of YSOs and
their disks \citep{Kul08}.

A correlation has been found between the soft X-ray and optical variability for disk bearing YSOs
in NGC 2264  \citep{Fla10}. 
This is thought to be due to absorption of both soft X-rays and optical photons by the dust disk.
It is possible that X-rays and/or X-ray variability are involved in some 
of the flux changes observed among YSOs in the IR. 
\citet{Ke12} argue that rapid aperiodic variations in the SEDs of  YSOs
could be induced by X-ray flares.
\citet{Fla12, Fla14} recently presented the first comprehensive study 
attempting to directly link X-ray and mid--IR behavior.  They monitored 
the young cluster IC 348 for 30 days in X-rays roughly overlapping their 40 
day IRAC monitoring period. They were specifically looking for reflex 
response of the disk to X-ray flares.  One could imagine that a flare in 
X-rays would lead to an increase in IR-brightness of the disk as the dust 
absorbed and re-radiated the X-ray energy. However,  \citet{Fla14}  found 
no evidence for a link between the X-ray and infrared variability on these 
timescales among 39 cluster members with circumstellar disks. They found 
no correlation between the shape of the X-ray lightcurves and the 
size of IR variability. Only weak X-ray flares were detected, and none 
correlated with a change in the infrared photometry on 
timescales of days to weeks following the flare. 

\subsection{Our Target GGD~12-15}

GGD~12-15, identified originally as a series of Herbig-Haro--like objects \citep{Gyu78}, is part of the Mon R2 star formation region\footnote{For a review, see \citet{Car08}.} at a distance of about 830$\pm$50~pc \citep{Rac68}. Mon R2 was first recognized as a chain of reflection nebulae which are now understood to be associated with one of the closest massive star forming regions to the Sun. The GGD~12-15 complex lies about 1 degree  west of the Mon~R2 core. Signatures of star formation activity from the region have been studied for over 30 years.  \citet{Rod80,Rod82} identified a water maser associated with a large bipolar molecular outflow. The molecular clump at the center of the region has a mass of $\sim280 M_\odot$ \citep{Lit90}.

An embedded infrared cluster has since been identified and studied in detail, including with  \textit{Spitzer} by \citet{Gut09}, who provided infrared YSO classifications for 119 sources in this region, encompassing 17 Class~I (star + disk+ envelope) and 102 Class~II (star + disk) sources. They demonstrated that the Class I YSOs are distributed in a non-symmetric pattern that still strongly resembles the distribution of gas and dust. In contrast, the presumed older Class II sources are distributed more widely. 
 Using infrared integral field spectroscopy, \citet{Maa11} studied the stellar content of the cluster, identifying two very young and massive B stars as well as several pre-main sequence stars of spectral types G and K. As a signpost of high-mass star formation, this region also hosts a bright cometary \ion{H}{2} region which is powered by an early B star \citep{Kur94,Gom98}. This region was subsequently shown to be associated with an expanding \ion{H}{1} region \citep{Gom10}.

In this paper, we present $Spitzer$ IRAC 3.6 and 4.5\micron\  lightcurves for YSOs and YSO candidates in the GGD~12--15 region. To this, we add a deep, contemporaneous/simultaneous X-ray observation from $Chandra$.  The X-ray observation serves two purposes. First, since young stars are known to be relatively bright X-ray sources, the X-ray data provide a sample of cluster members independent of their IR characteristics, and therefore the X-ray data allow us to include diskless YSOs in the study.  Secondly, we can test suggestions that X-ray activity is related to IR variability.

In the next section, we will discuss the general IR and X-ray observations and the bulk metrics used including luminosity and color and their observed changes. In \S 3, we summarize the bulk IR variability statistics and the basic X-ray results. We then examine the periodic sources and  how the IR variability is dependent on spectral energy distribution (SED) class in \S 4.  We discuss the importance of class on variability   and whether X-rays flux holds any hint to the IR behaviors as well as looking at GGD~12-15 in the context of other clusters in  \S 5. We summarize the work in \S 6.

\section{Data Reduction}
\label{sec:DR}
\subsection{Chandra Observations and Analysis}
\label{sec:DR_CXO}
\begin{figure*}[htbp]
\begin{center}
\includegraphics[width=6.5in]{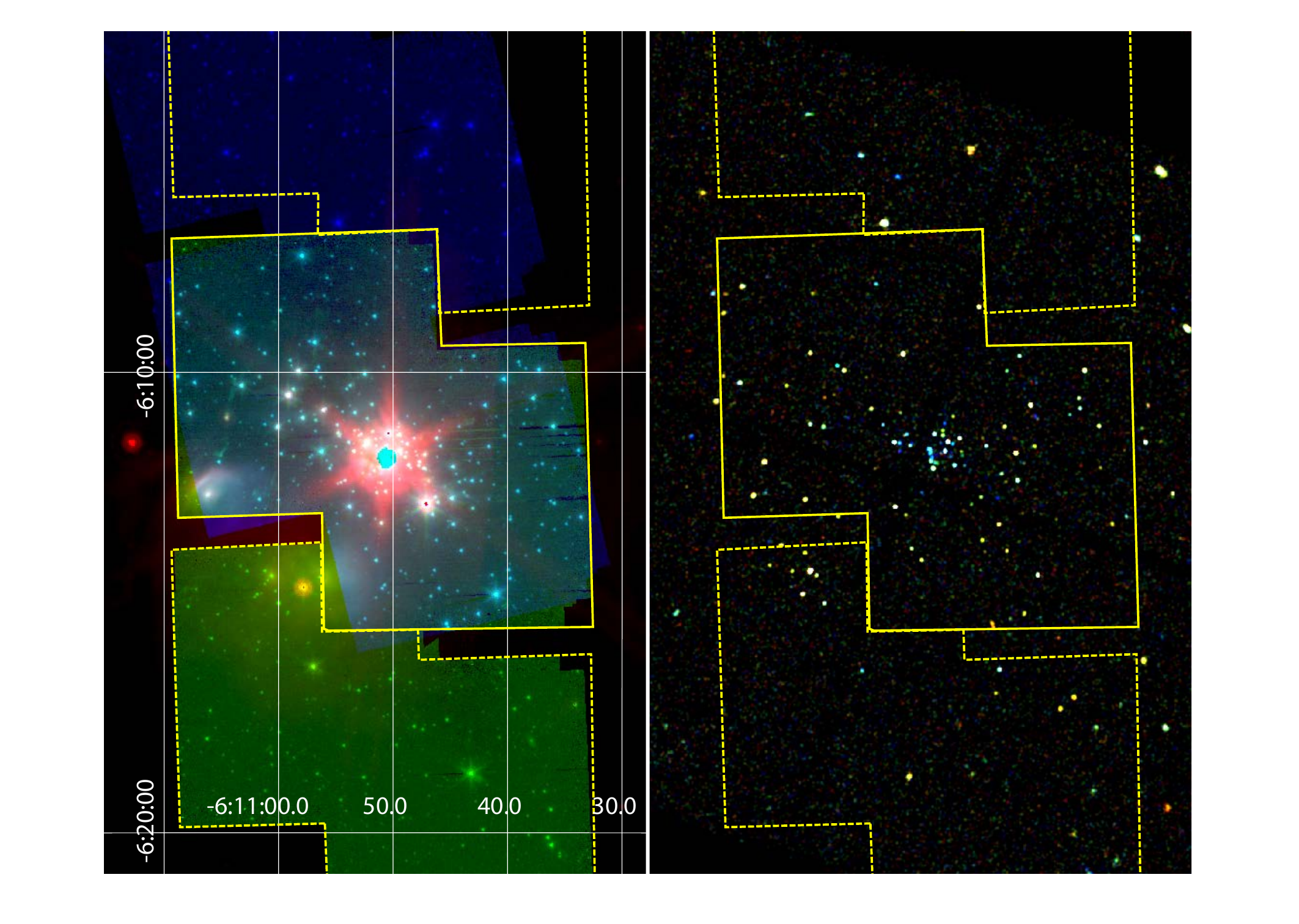}

\caption{Left: One epoch of \textit{Spitzer} data; IRAC-1 ([3.6]), IRAC-2 ([4.5]) and MIPS-24 (blue, green and red respectively).  The solid yellow polygon represents the overlap region, the dotted yellow polygons indicate typical 1 channel regions for the YSOVAR campaign  -- IRAC-2 is towards the South and IRAC-1 is towards the North.   Right: The $Chandra$ data, Red: energy $<$ 1.5 keV, Green: 1.0$<$ energy $< $2.5 keV, Blue: 2.1 $<$ energy $<$ 4.0 keV.  Yellow polygons are taken from the left frame.   The overlap region for a single epoch is about 104 sq arc minutes (out of 289 sq arc minutes on ACIS-I). Rotation of the $Spitzer$ field over the course of the monitoring period brings the total overlap to about 150 sq arc minutes, or about half the ACIS field.}
\label{visuals}
\end{center}
\end{figure*}

GGD~12-15 was observed with the $Chandra$ X-ray Observatory (CXO) 15 December 2010 for about 67 ks, starting at 14:04 UT (ObsID~12392).
Because of field rotation over the 38 days of the $Spitzer$ observations, the band coverage and duration of the $Spitzer$ lightcurves for each source depends sensitively on the exact position in the ACIS field of view. 
The aimpoint of the observation was 6:10:50, -06:12:00 (J2000.) at a roll angle of 18 degrees. The Advanced CCD Imaging Spectrograph (ACIS) was in the nominal imaging configuration (chips I0-I3) which provides a field of view of approximately 17\arcmin $\times$ 17\arcmin.  Data were taken in very faint mode to aid in the filtering of background events. 
In addition, the S2 and S3 chips were active, however the analysis of these data are not presented here. The location of the relevant portion of the X-ray field is divided into three parts, the northern part is dominated by [3.6] observations, the southern part is dominated by [4.5] $Spitzer$ data, while we refer to the central region as the ``overlap'' regions wherein most stars are well observed in both channels.    Total rotation of the \textit{Spitzer} field changes  by about 20 degrees during the course of the monitoring. The result is that many sources to the north have only some [4.5] observations and vice-versa, while coverage in the overlap region is not perfect for all sources in both bands. Typical data are shown in Figure~\ref{visuals}.

The data used in this analysis were processed with $Chandra$ X-ray Center Data System version 8.4. 
As such, they were processed through the standard $Chandra$ Interactive Analysis of Observations \citep[CIAO;][]{Fru06} pipeline at the
$Chandra$ X-ray Center. This version of the pipeline automatically employs a sub-pixel positioning algorithm, charge-transfer inefficiency correction, and  incorporates a noise correction for low energy events. This last correction can remove good events
from the cores of bright point sources, resulting in an underestimation of the X-ray flux.
In this case, the count rate did not exceed 0.02 counts per second for any source, thus event loss is not a concern.
Background was nominal and non-variable. 

To identify point sources, photons with energies below 300 eV and above 8.0 keV were filtered out from this merged event list. This excludes energies that generally lack a stellar contribution.  By filtering the data as described, contributions from hard, non-stellar sources such as X-ray binaries and active galactic nuclei (AGNs) are attenuated, as is noise. A monochromatic exposure map
was generated in the standard way using an energy of 1.5 keV, which is a reasonable match to the expected peak energy of the stellar sources and the $Chandra$ mirror transmission. The CIAO tool WavDetect was then run on a series of flux-corrected images binned by 1, 2, and 4 pixels.  The output source lists were combined and this resulted in the detection of 164 sources with greater than 3 $\sigma$ source significance. 

At this level of significance, only 1 false detection is expected; however, not all sources are expected to be YSOs associated with GGD~12-15.  Using the extragalactic luminosity relationship for $Chandra$ \citep[log$N$-log $S$;][]{Gia01}, we estimate about 60 extragalactic sources -- most of these will be indicated by the lack of any counterpart in the Two Micron All Sky Survey \citep[2MASS][]{Skr06} point source catalog.
Other stellar contamination includes active sources along the line of sight.  Estimates from \citet[IC1396;][]{Get12} and \citet[RCW 38, RCW 108;][]{Wol06, Wol08} indicate that there should be about 20 foreground objects. Overall, these background and foreground numbers are consistent with about 90 X-ray sources associated with GGD~12-15.   

\skipthis{As was described in Paper~I, we include source detections by the WavDetect algorithm at low levels of significance.   We identify 210 candidate sources in the regime of source significance between 2 and 3 (similar to 2--3 $\sigma$ detections).  We emphasize that many of these are not X-ray sources but just background fluctuations along with dozens of very weak extragalactic sources and a few stellar sources that can be identified via their bright IR counterparts.
This relatively low significance admits for the possibility of several false positives.  
But in this case, we are only concerned with YSOs in the IRAC field and all YSOs should have relatively bright counterparts among the IRAC sources.   This is not to limit the sample to stars with disks, but to limit the sample to stars with photosphere bright enough to be cluster members(see \S \ref{sec:DR_SST}). }

As was described in Paper~I, we include source detections by the WavDetect algorithm at low levels of significance.   We identify 210 candidate sources in the regime of source significance between 2 and 3 (similar to 2--3 $\sigma$ detections).  We emphasize that many of these are not X-ray sources but just background fluctuations along with dozens of very weak extragalactic sources and a few ($\sim$ 15) stellar sources that can be identified via their bright IR counterparts.  This relatively low significance admits for the possibility of several false positives.  But in this case, we are only concerned with YSOs in the IRAC field and all YSOs should have relatively bright counterparts among the IRAC sources (see \S \ref{sec:DR_SST}).

For each $Chandra$ source, net counts and energy distributions were calculated as well as a timing index. 
To calculate net counts, a background ellipse is identified. The background is an annular ellipse with the same center, eccentricity, and rotation as the source.  The outer radius is 6 times the radius of the point spread function (PSF) at the off-axis distance of the source. The inner radius is 3 times larger than the PSF. From this region, any nearby sources are subtracted with ellipses 3 times the size of the source ellipse. 
The net counts are calculated by subtracting the background counts (corrected for area) and multiplying the result by 1.053 to correct for the use of a 95\% encircled energy radius \citep{Wol06}. 

For each source, a lightcurve is generated and its Gregory--Loredo variability statistic \citep[GL-vary;][]{Gre92} is calculated to characterize the source variability. This method uses maximum-likelihood statistics and evaluates a large number of possible break points from the prediction of constancy. GL-vary is able to evaluate the probability that the source was variable as well as estimate the constant intervals within the observing window.  GL-vary returns an index parameter that addresses the degree of the variability. The higher the value of this index, the greater the variability; values greater than 7 indicate $>$99\% variability probability.  A GL-vary index above 8 usually indicates a flare.  Both the GL-vary index as well as spectral fits, prove unreliable below 30 net counts. For sources with over 30 net counts,  a spectrum is fitted using  the Astrophysical Plasma Emission Code (APEC) model of a one--temperature plasma with gas absorption\citep{Smi01}. YSOs are found to be reasonably described by such a plasma. We assumed 0.3 times solar elemental abundances previously suggested as typical for YSOs in other star-forming regions \citep{Ima01, Fei02, Get05}. 

\subsection{Spitzer Observations and Analysis}
\label{sec:DR_SST}

As part of YSOVAR, the field was observed using $Spitzer$ from 16 November to 24 December 2010 -- these data are referred to as ``YSOVAR'' data.  The YSOVAR data were generated by a pair of pointings offset by 2.5\arcmin\ in both RA and Dec with a field center of 6:10:48, -06:12:30 (J2000.; See Fig.\ \ref{visuals} \& Paper~I; Fig.\ 5).  
Fast cadence was used, meaning the observations were made in a series of 3.5 day cycles.  Within each cycle, the time step between visits was about 4, 6, 8, 10, 12, 14 and 16 hours.  By using a linearly increasing time-step, we were able to evenly sample, in Fourier space, a range of higher frequency (shorter timescale) photometric variability while minimizing the total amount of necessary observing time to sample these high frequencies and to minimize the period aliasing (see Paper~I).

The YSOVAR data consist of 77 observations of GGD 12--15 over a 38 day period with a total exposure time of 14.7 hours. 
The orientation of the IRAC field of view on the sky depends on ecliptic latitude and date of observation. This results in a rotation of the field of view of about 20 degrees over the 38 day campaign.  This is mitigated somewhat by the use of two pointings.  There was also a 20 hour staring mode observation which overlapped the X-ray observation, from 14:06 UT  Dec 15 to 09:39 UT Dec 16, 2010.   Those data began and ended with regular maps in the two IRAC channels that are included in the current study - bringing the maximum number of observations of any star to 79.   The staring mode data themselves will be part of a separate study and are not discussed here. 

A detailed account of the data processing and the source extraction is given in paper I. Here we summarize the main data reduction and processing steps.  Basic calibrated data (BCD) are obtained from the \textit{Spitzer} archive. Further data reduction is performed with the Interactive Data Language (IDL) package cluster grinder \citep{Gut09}, which treats each BCD image for bright source artifacts. Aperture photometry is run on individual BCDs with an aperture radius of 2.4\arcsec. To increase the signal-to-noise ratio and reject cosmic rays,   the photometry from all BCDs in each visit is combined. The reported value is the average brightness of all the BCDs within that visit which contains the source in question, after rejecting outliers. The photometric uncertainties obtained from the aperture photometry are, particularly for bright sources, only lower limits to the total uncertainty, since distributed nebulosity is often found in star forming regions and can contribute to the noise. To improve these estimates, Paper~I introduces an error floor value that is added in quadrature to the uncertainties of the individual photometric points. The value of the error floor is 0.01 mag for [3.6] and 0.007 mag for [4.5].

Following Paper~I, we retain all sources that have at least 5 data points in either [3.6] or [4.5]. We also require a detection to have been reported by \citet{Gut09}, this requirement cut down on weak sources near strong ones which might be detected inconsistently. We visually inspected all frames for lightcurves that are classified as variables in \S\ref{sec:Results} and removed data points visibly affected by instrumental artifacts.  Embedded star-forming cores like GGD~12-15 pose significant problems for deriving accurate light curves with Spitzer/IRAC because of crowding issues, field rotation during the campaign causing column pull-down effects and scattered light effects to migrate over the cluster image, latent image effects, and pixel-phase effects.  We have done our best to provide good light curves by the design of our observations, and by inspecting and cleaning all of the light curves discussed in this paper.   However, some artifacts may remain which may, for example, influence some of the CMD slopes we derive.   We believe that our conclusions are nevertheless robust to any remaining artifacts in the data.

\begin{figure*}[htbp]
\begin{center}
\includegraphics[height=4.5in]{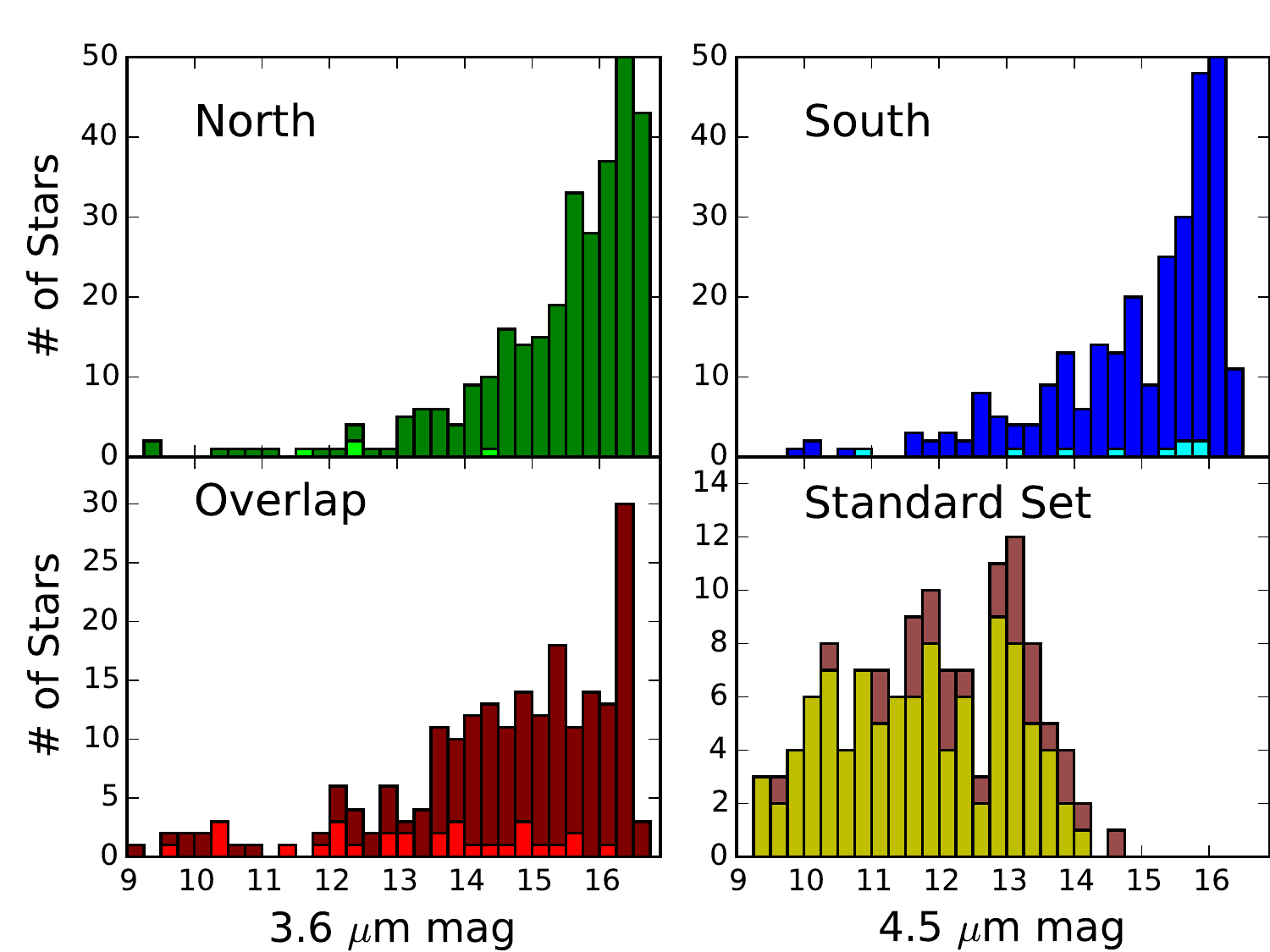}

\caption{Histograms of the magnitude distributions of objects in the northern (top-left), overlap (bottom-left) and southern (top-right) regions as well as the standard set (bottom-right).  The standard set stars have been removed from the other three samples. The vertical axis indicates the number of stars in each 0.25 magnitude bin.  Variables are indicated by the lighter color. Notice that the overlap field has far more and brighter variables than the northern or southern regions -- even in the faintest magnitude bins. 
The different X-axes are due to the fact that the northern set was mostly observed at [3.6] and the southern set mostly in [4.5].
}
\label{depth_hist}
\end{center}
\end{figure*}

Overall, 1017 IR sources were analyzed. 
Of these, about 1/3 were in the field observed by both detectors.  Figure~\ref{depth_hist} shows a histogram of the magnitude distribution found in the three fields. The central field, in which the two detectors overlapped, has a distinctly flatter distribution than either the northern or southern fields.  This may be an indication of crowding in the central region resulting in a higher background, making good photometry more difficult for the fainter stars compared to the the less crowded northern and southern fields. Furthermore, both the northern and southern fields have only a small number of variables fainter than $\sim 14^{th}$ mag. We take this to indicate that the YSOs are highly concentrated and mostly lie in the overlap region (see e.g., Carpenter \e 2000).  In Paper~I, we also defined a ``standard set'' of members for all tests so that direct comparison of results among the clusters studied in the YSOVAR project would be possible.   The ``standard set'' of members includes all stars identified by \citet{Gut09} as YSOs.   This is the majority of disked members. We then add to this list all X-ray sources that have a brightness and  SED slope (based on all available data) appropriate for a Class III (diskless) star at the distance of GGD~12-15.  Variables are identified three ways, via Stetson index \citep{Ste96}, high $\chi^2$ values when compared to a constant source or strong periodicity (see \S~\ref{sec:IRVar} for details).    

Following Paper~I and \citet{Gun14}, sources are cross-matched from individual observations with a matching radius of 1\arcsec\ with each other and the 2MASS catalog, which is used as a coordinate reference. All photometric measurements performed in the context of the YSOVAR project are collected in a central database which we intend to deliver to the Infrared Science Archive (IRSA)\footnote{http://irsa.ipac.caltech.edu/} for general distribution. Data for this article were retrieved from our database on 2014--09--23 and further processed using custom routines in Python.\footnote{Code is available at https://github.com/YSOVAR.}  Table~\ref{TAB:IR} contains the basic IR data  for the stars and their lightcurves. Some of the basic parameters tabulated include:
\begin{itemize} 
\item Identification of the star as part of the standard set of cluster members and/or an X-ray source (see Paper~I).
\item Identification of the star as a variable and the related variability metrics.
\item The number of observations in each channel, the mean, median, standard deviation and median absolute deviation of the sampled lightcurve for each source. 
\item The width of the distribution of magnitude values in the lightcurve from the 10\% to the 90\% quantile. We defined this as $\Delta$.
\item  Parameters from a linear fit to the distribution of the points in color-magnitude space. We use an orthogonal distance regression
method that takes the errors in both the x and y directions into account on the data within the 10-90\% brightness quantiles \citep{Bog92}. We measure the angle of best fit and the length of the vector in magnitudes, and the 1 $\sigma$ errors on each. 
\end{itemize}

\begin{deluxetable}{ccccc}
\tabletypesize{\footnotesize}
\tablecaption{\label{TAB:IR} Source designations, flux densities and lightcurve properties.}
\tablehead{\colhead{Column ID} & \colhead{Name} & \colhead{Unit} & \colhead{Channel} & \colhead{Comment}}
\startdata
1 & IAU\_NAME & -- & -- & J2000.0 IAU designation. \\ 
2 & RA & deg & -- & J2000.0 Right ascension. \\
3 & DEC & deg & -- & J2000.0 Declination. \\
4 & n\_36 & ct & $3.6\;\mu$m & Number of datapoints in the lightcurve at  [3.6].\\
5 & n\_45 & ct & $4.5\;\mu$m & Number of datapoints in the lightcurve at [4.5].\\
6& n\_3645 & ct &$3.6\;\mu$m, $4.5\;\mu$m & Number of datapoints in the lightcurve with both colors.\\
7& mean\_36 & mag & $3.6\;\mu$m & Mean mag of lightcurve. \\
8& mean\_45 & mag & $4.5\;\mu$m & Mean  mag of lightcurve. \\
9 & median\_36 & mag & $3.6\;\mu$m & Median mag.\\
10 & median\_45 & mag & $4.5\;\mu$m & Median mag.\\
11 & mad\_36 & mag & $3.6\;\mu$m & Median absolute deviation of the  lightcurve. \\
12 & mad\_45 & mag & $4.5\;\mu$m &  Median absolute deviation of the lightcurve. \\
13 & delta\_36 & mag & $3.6\;\mu$m & Width of distribution from 10\% to 90\%. \\
14 & delta\_45 & mag & $4.5\;\mu$m & Width of distribution from 10\% to 90\%. \\
15& redchi2tomean\_36 & -- & $3.6\;\mu$m & Reduced $\chi^2$ to mean. \\
16 & redchi2tomean\_45 & -- & $4.5\;\mu$m & Reduced $\chi^2$ to mean. \\
17 & coherence\_time\_36 & days & $3.6\;\mu$m & Decay time of ACF \\
18 & coherence\_time\_45 & days & $4.5\;\mu$m & Decay time of ACF \\
19 & stetson\_36\_45 & -- & $3.6\;\mu$m, $4.5\;\mu$m & Stetson index for a two-band lightcurve. \\
20 & cmd\_length\_36\_45 & mag & $3.6\;\mu$m, $4.5\;\mu$m & Length of best-fit line in [3.6], [3.6]-[4.5] CMD. \\
21 & cmd\_alpha\_36\_45 & rad & $3.6\;\mu$m, $4.5\;\mu$m & Angle of best-fit line in  [3.6], [3.6]-[4.5] CMD.\\
22 & cmd\_alpha\_error\_36\_45 & rad & $3.6\;\mu$m, $4.5\;\mu$m & 1 $\sigma$ error on angle in  [3.6], [3.6]-[4.5]CMD.\\
23& Jmag & mag & $J$ & 2MASS J mag.  \\
24& Hmag & mag & $H$ & 2MASS H mag.\\
25& Kmag & mag & $K$ & 2MASS K mag. \\
26 & 3.6mag & mag & $3.6\;\mu$m & Cryo [3.6]. \\
27 & 4.5mag & mag & $4.5\;\mu$m & Cryo [4.5].  \\
28 & 5.8mag & mag & $5.8\;\mu$m & Cryo [5.8]. \\
29 & 8.0mag & mag & $8.0\;\mu$m & Cryo [8.0]. \\
30 & 24mag & mag & $24\;\mu$m & Cryo [24].\\
31 & e\_Jmag & mag & $J$ & Observational uncertainty. \\
32& e\_Hmag & mag & $H$ & Observational uncertainty. \\
33& e\_Kmag & mag & $K$ & Observational uncertainty. \\
34 & e\_3.6mag & mag & $3.6\;\mu$m & Observational uncertainty (cryo data). \\
35 & e\_4.5mag & mag & $4.5\;\mu$m & Observational uncertainty (cryo data). \\
36 & e\_5.8mag & mag & $5.8\;\mu$m & Observational uncertainty. \\
37 & e\_8.0mag & mag & $8.0\;\mu$m & Observational uncertainty.\\
38 & e\_24mag & mag & $24\;\mu$m & Observational uncertainty. \\
39 & IRclass & -- & -- & IR class according to \citet{Gut09}. \\
40 & SEDclass & -- & -- & IR class according to SED slope.  \\
41 & Variable & binary &1= yes &  Source is determined to be a variable.\\
42 & X-ray &  binary &1= yes  &  $Chandra$ counterpart.\\
43 & Standard & binary &1= yes &  Source in YSOVAR standard set of cluster members.\\
44 & Aug & binary &1= yes  &  In the augmentation to the standard set of members.\\

\enddata
\tablecomments{Columns 26-39 are taken directly from \citet{Gut09} and republished here for completeness. This table is published in its entirety in the electronic version of the journal. Here the table columns are described as a guide to form and content.}
\end{deluxetable}

\section{Results}
\label{sec:Results}

\begin{figure*}[htbp]
\begin{center}
\includegraphics[width=5.5in]{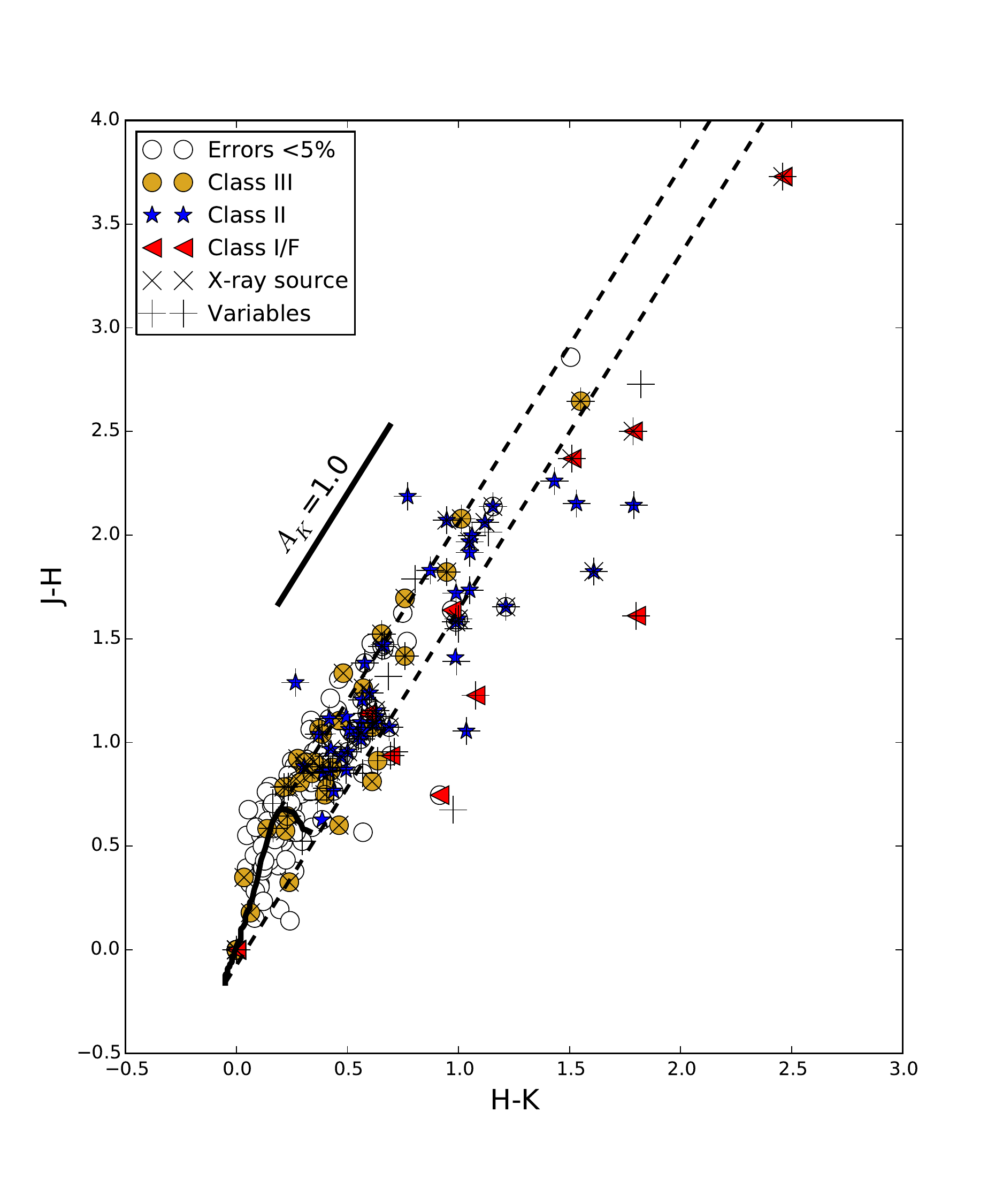}
\caption{A near-IR two-color diagram for GGD~12-15 based on colors from 2MASS. 
The red triangles indicate SED class Class~I or flat, while Class~II and III objects are represented by blue stars and gold circles respectively.
X-ray sources are marked with ``X" and $Spitzer$ variables are marked with ``+".
The open circles indicate all sources with 2MASS errors $<5\%$ in all three channels which have lightcurves, but are not in the standard set.
The curved black line indicates the main sequence; the parallel dashed lines indicate the direction of the reddening vector bracketing the main sequence.  The short solid line indicates $A_K=1.0$.  Most sources in this plot show reddening $<0.5 A_K$, but there are some more extreme cases among the Class II, flat spectrum and Class I objects.}
\label{NIR_TDC}
\end{center}
\end{figure*}

GGD~12-15 was briefly discussed in a cluster survey paper by \citet{Gut09} which combined 2MASS and $Spitzer$ photometry of stars in several clusters to identify YSOs and the structure of the individual clusters.  However,  2MASS data are not very deep and are incomplete for YSOs in GGD~12-15. Figure~\ref{NIR_TDC} shows all the stars in the field with  2MASS photometric errors of $<$5\%.  
There are 74 standard set members detected in 2MASS with errors $<$5\%; of these, 49 vary in the $Spitzer$ bands. 
The 2MASS standard set members break down as 16 Class I/F candidates, 32 Class II candidates and 26 Class III candidates.  
  The median extinction of the 2MASS sources is about 0.5  A$_K$, but there are many sources with A$_K$ $>$ 1.
\citet{Gut09} list the exposure time of the $Spitzer$ cryo data as 41.6 seconds in all 4 IRAC passbands and about 40 seconds in the MIPS 24~\micron\ channel.  This leads to a 90\% differential completeness of 16.3, 16.1, 14.3, 12.9 and 8.4 in the 5 respective channels \citep{Gut09}.

\begin{figure*}[htbp]
\begin{center}
\includegraphics[width=5.5in]{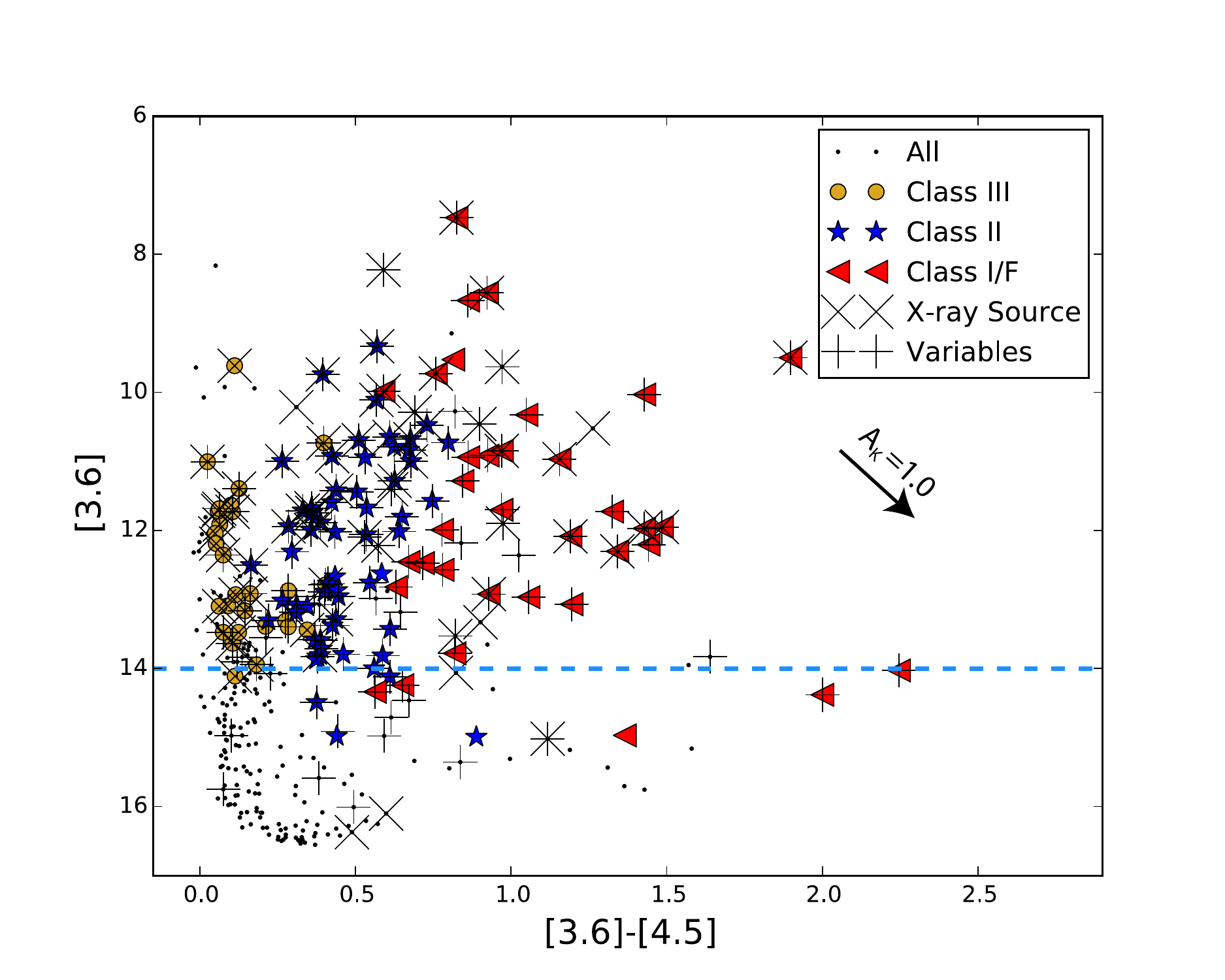}
\caption{A color-magnitude diagram based on colors from the mean of the YSOVAR data. 
The red triangles, blue stars and gold circles are as in the previous figure.
 X-ray sources are marked with ``X" and variables are marked with ``+".
Small dots indicate all sources which were detected in the overlap field at least 5 times.
The horizontal dashed blue line indicates the brown dwarf limit in the absence of reddening  \citep{Sie00} .
The $A_K=1.0$ reddening vector is indicated \citep{Ind05}. 
}
\label{MIR_CMD}
\end{center}
\end{figure*}

Spitzer [3.6] and [4.5] data should be complete with respect to unabsorbed stars in the GGD~12-15 cluster.
Mass tracks from \citet{Sie00} indicate that a 0.1 $M_\odot$ star at 1 Myr should have an absolute magnitude of about 4 at 3.6\micron.  Given the distance modulus of 9.6 \citep{Car08}, most cluster members should be brighter than [3.6]$\approx$14.   However as seen in Figure~\ref{MIR_CMD}, there are at least some YSOs fainter than [3.6] =14, evidence there is significant reddening for many of the flat spectrum and Class~I objects.  This figure indicates about 100 disked objects. 

\skipthis{We identify high quality data in the shorter two channels as having errors less than 5\% in both channels; there are about 375 of these.  
High quality data in the longer channels is defined as being high quality in the shorter wavelengths and as having error $<$10\% in the longer wavelengths there are about 160 of these. 
Even in the deepest of these subsets shows fairly mild extinction with median A$_K \sim$ 0.25.  
}

The issue of the disk status or ``Class'' is subtle.  
The scheme proposed by \citet{Gre94} uses the shape of the mid-IR SED of candidate YSOs to sort them into different classes
ranging from photospheric through stars with disks to those dominated by the infalling envelope.
The SED of the observed star is compared to that expected from a normal photosphere. If excess infrared emission is seen in the target star when compared to a normal photosphere, this excess can be interpreted as emission from a disk. In practice, several things complicate the identification of young stellar objects from infrared colors alone. First of all, galaxies with high rates of star formation can mimic the appearance of a single star with an envelope--like infrared excess.  These galaxies tend to be fainter than most member stars. Secondly,  infrared disks can have differing amounts of excesses associated with them. Face-on disks look different from edge-on disks  \citep{Mey97}, some can have cleared centers, and others can change accretion rate  \citep{Ric12}.  Finally, extinction along the line of sight can flatten the slope measured on the SED.

Because of ambiguities associating colors and disk class,  \citet{Gut09} chose a very conservative system of magnitude and color cuts for identifying sources as being Class I or II.   \citet{Gut09} cannot comment on many sources that simply lack data in certain bands. 
We use this color-based classification to identify the star as part of the standard set of cluster members.  However, when comparing classes we use the SED slope, which can be measured for all members of the survey.  


\subsection{Basic IR Variability Statistics of GGD~12-15}
\label{sec:IRVar}

We ran several statistical tests on the distribution of the photometric data for each source.  These have been discussed exhaustively in Paper~I and revisited in detail by \citet{Gun14}, hence we only quickly touch on key statistics below. For all sources, the mean and median magnitude were calculated.  We also calculated the  median absolute deviation,  as well as the reduced $\chi^2$ to the mean of the magnitude and the magnitude range between the 10\% and 90\% quantiles as initial (outlier resilient)  measurements of the variability of each source.  The reduced $\chi^2$, while very well understood, is not very sensitive to low amplitude variability. Instrumental uncertainties lead to a non-Gaussian error distribution, and we are forced to use a conservative cutoff and select variable sources only if $\chi^2$ was greater than 5 (see Paper~I).
If the source is observed in both channels, the Stetson index is a more robust variability index \citep{Ste96}.   The Stetson index $S$ is a fairly well-known test for variability when the variability is correlated in two or more channels. This is well discussed in the literature \citep[e.g.\ Paper~I;][]{Gun14,Ric12,Car01}. A larger value of the $S$ index indicates larger coherent variability -- the exact dividing line between variable and not variable requires careful analysis of each source, since the number of samples and errors on each observation enter into the precise determination. However, following Paper~I, we consider sources with $S>$ 0.9 to be undoubtably variable.

Generally, variable YSOs do not move randomly in color-color or color-magnitude space.
Instead, they trace out a trajectory which is often linear in the NIR  \citep{Car01,Ric12}. For example, \citet{Ric12} found many stars which trace out a line which is parallel to the classical T Tauri star locus  \citep{Mey97} in the $J-H$, $H-K$ two color diagram.   Since we monitor in two channels, we can examine the color-magnitude space.  We find that the points occupied by the individual samples a single star can be fitted to a slope which is well defined in many of the [3.6] vs. [3.6]-[4.5] color-magnitude diagrams (CMDs).   For each source, we fitted a straight line to all points in the CMD using an orthogonal distance regression method that takes the errors in both the color and the magnitude into account. This allows us to compare, in a quantifiable way, the trajectory that the variable YSOs follow in color-magnitude space. We will discuss the results and meaning of these fits in detail in \S 5.1.  

 Following Paper~I, we performed a period search on all lightcurves with at least 20 points.  
We computed results using several methods: Lomb-Scargle \citep [LS;][]{Sca82}, box-fitting least squares \citep{Kov02}, Plavchan \citep{Pla08} as well as  the autocorrelation function (ACF) on the  3.6\micron, 4.5\micron\ and, where possible, the [3.6]$-$[4.5] lightcurves.  Given the overall
sampling of our data, we require at least two and a half periods over the window of  the collective YSOVAR observations, so we looked for periods between only 0.1 and 14.5 days. Typically, if the LS algorithm found a reliable period, the other approaches found comparable periods and this period was used. If the results were inconsistent, no period is given. 

We dropped candidate periodic objects if the calculated false alarm probability (FAP) was $>$ 0.03. 
 For each of the remaining objects, we investigated the phased lightcurve. 
In priority order, we take any period derived from the [3.6] data first ([3.6] is less noisy than [4.5]), then, only if there is no [3.6] period of sufficient power, we take the period derived from [4.5].  We also derived a period from the [3.6]$-$[4.5] color. However in GGD~12-15, no periods were derived on the basis of being periodic in color alone. 

\begin{figure*}[htbp]
\begin{center}
\includegraphics[width=6.5in]{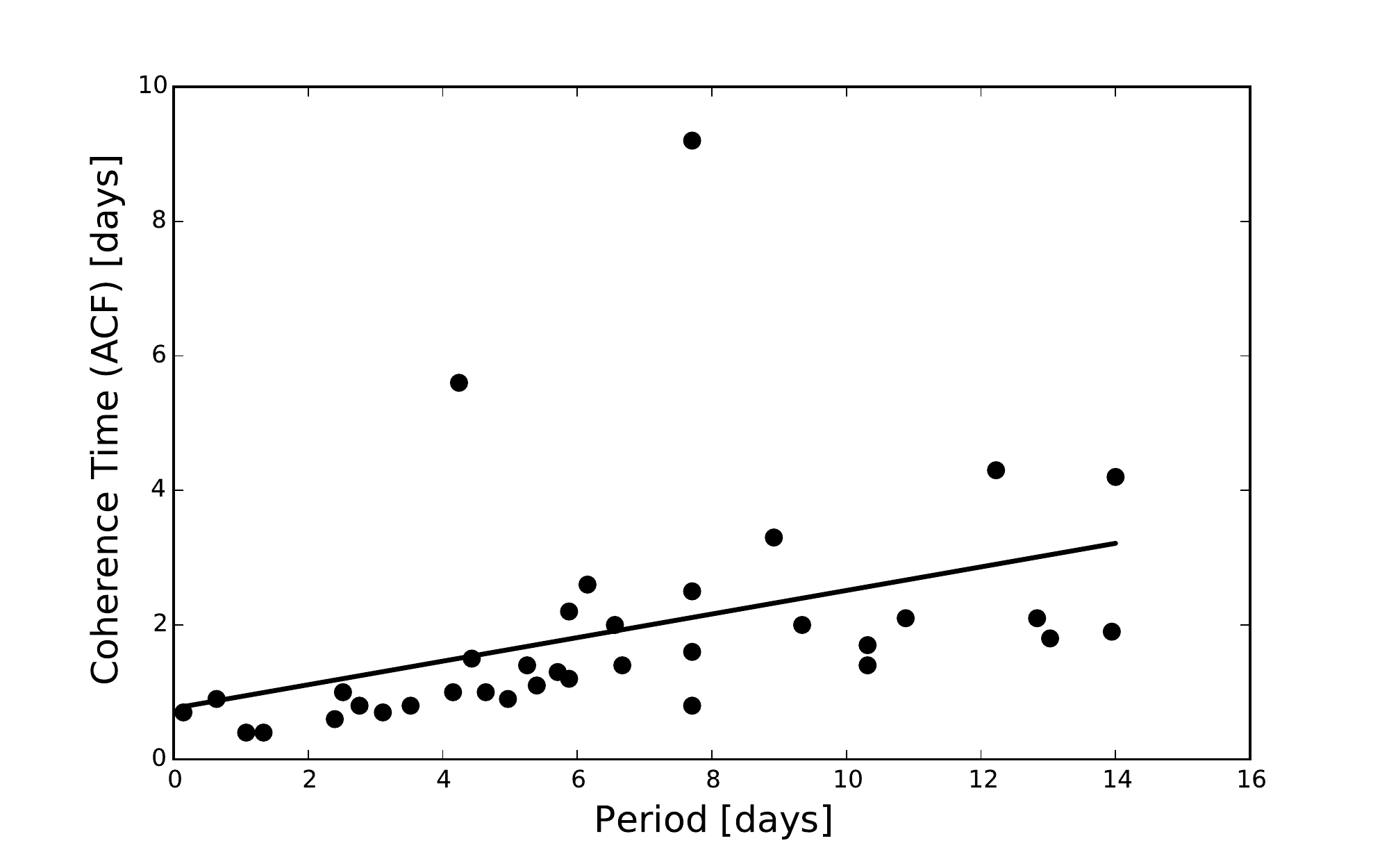}

\caption{
Plot of 3.6 \micron\ coherence time versus period of all sources with both measurements.  The line is a fit to all the data  using a linear regression (not weighted by errors which are formally very small).  The fit is clearly offset due to two outliers with relatively high coherence time (or low variability). The $\sim$ 5:1 slope ratio found for the line appears consistent with the bulk of the periodic sources. 
}
\label{periodVACF}
\end{center}
\end{figure*}

As mentioned above, the discrete auto-correlation function (ACF) of each lightcurve is calculated. The ACF is statistically well defined for evenly sampled data: it is the cross correlation of a function with itself. When the starting times ($\tau_0$) are the same, the correlation coefficient is one. As the $\tau$ values become different, the correlation coefficient drops. For a periodic signal, there would be a minimum with a negative correlation at the time of the half period and a secondary maximum at the time scale of the period. Non-periodic data tend toward zero correlation. Following \citet{Gun14}, we take the first value of $\tau$ with ACF($\tau$) $<$ 0.25 as the {\it coherence time} for the light curve.\footnote{A more common definition is to use the position of the first local maximum in the ACF, but due to the low number of data points in our lightcurves, the noise in the ACF is large and this value is often not well defined.}  In Fig.~\ref{periodVACF} we plot the periods of the periodic sources in GGD~12-15  on the X-axis versus the time (in days) at which the ACF value dropped below 0.25 for those objects.  The values appear linearly correlated.  Specifically,  using a linear regression which included the two outliers, we find the coherence time in [3.6] is about 20\% of the period.   \citet{Pop15} find a similar result in IRAS 20050+2720 -- with a coherence time of about 30\% of the period. The significance of this discrepancy (20\% vs. 30\%) is still under investigation. We do not mean to imply that the coherence time gives us the timescale of the physical processes at work; it provides for relative comparison of long versus short timescales for changes in the non-periodic YSOs.

\subsection{X-Ray Results}
\label{sec:XrayResults}
There are 164 
 X-ray sources detected at 3$\sigma$ or greater significance. This corresponds to $<$ 1\% false detection probability for WavDetect source detection \citep{Fre96}.  
There were an additional  $\sim$ 210 
X-ray sources  detected at a significance of between 2 --3 $\sigma$ (total of $\sim$380).  Lacking additional data, these sources would typically be ignored.  However, the 2MASS and IRAC images are supporting data and we accepted as real X-ray sources those with source significance between 2--3 which were matched to IRAC sources (see \S\ref{sec:DR_SST}).  
Random matches with $Spitzer$ sources are not likely, even for weak X-ray sources.  There were only six IR sources matched to any of the low confidence (2 --3 $\sigma$) X-ray sources. The matching area is 1.41 sq. arcsec. per source $\times$  210 X-ray sources = 296 sq. arc sec.  The  total area is 1,040,400 sq arc sec.  So the fractional matching area filled by 2-3 $\sigma$ X-ray sources  is 0.0284\% 
of ACIS (\ie1/3500 chance of a random match) since there are about 1000 IR sources - total, the probability is for 0.2-0.3 random matches.   In all, 90 X-ray sources had IR counterparts within 1\arcsec. Forty-seven of these were bright enough for spectral calculation. 
The X-ray characteristics of the sources with IR counterparts are given in Table~\ref{Tab:Xraysum}. 

Nine X-ray sources were found to be variable at greater than 99\% confidence (using the GL-vary criteria from \S 2.1).  Of these,  only SSTYSV J061101.19-061417.7 was noted to flare strongly in X-rays. Unfortunately, this source is near the edge of the monitoring field with only 8 detections in the [4.5] field and none at [3.6]. It is identified as Class III based on previously acquired mid- and near-IR data. 

\begin{deluxetable}{lccr}
\tabletypesize{\small}
\tablecaption{\label{Tab:Xraysum} Properties of X-ray sources.}
\tablehead{\colhead{ID} & \colhead{Name} & \colhead{Unit} & \colhead{Comment}}
\startdata
1 & IAU\_NAME & J2000.0&  IAU designation of matched IR source within the YSOVAR program \\
2 & RA & deg &  J2000.0 Right ascension of X-ray source \\
3 & DEC & deg &  J2000.0 Declination of X-ray source\\
4 & Raw Counts & -- &  Raw X-ray counts \\
5 & Net Counts & -- &   Net X-ray counts\\
6 & N$_H$   & 10$^{21}$ cm$^{-2}$ & Fitted N$_H$ column \\
7 & N$_H$ err.  & 10$^{21}$ cm$^{-2}$ &  Error in fitted N$_H$ column \\
8 & kT& keV &  Fitted temperature \\
9 & kT err.  & keV & Error in fitted temperature \\
10 &Absorbed Flux& erg cm$^{-2}$ sec$^{-1}$ &  Flux from source after correcting for absorption. \\
11 & Unabsorbed Flux  & erg cm$^{-2}$ sec$^{-1}$  &   Flux from source without correcting for absorption. \\
12 & $\chi^2$ & -- &  Reduced $\chi^2$ of the spectral fit \\
13 & Statistic& --& c-statistic or $\chi$-data variance used?\\
13 & G-L Vary prob & -- & Probability that the source is an X-ray variable\\
14 & G-L Vary index & -- & G-L vary index. \\
\enddata
\tablecomments{This table is published in its entirety in the electronic version of the journal. Here the table columns are described as a guide to form and content.}
\end{deluxetable}

\begin{figure*}[htbp]
\begin{center}
\includegraphics[width=6.5in]{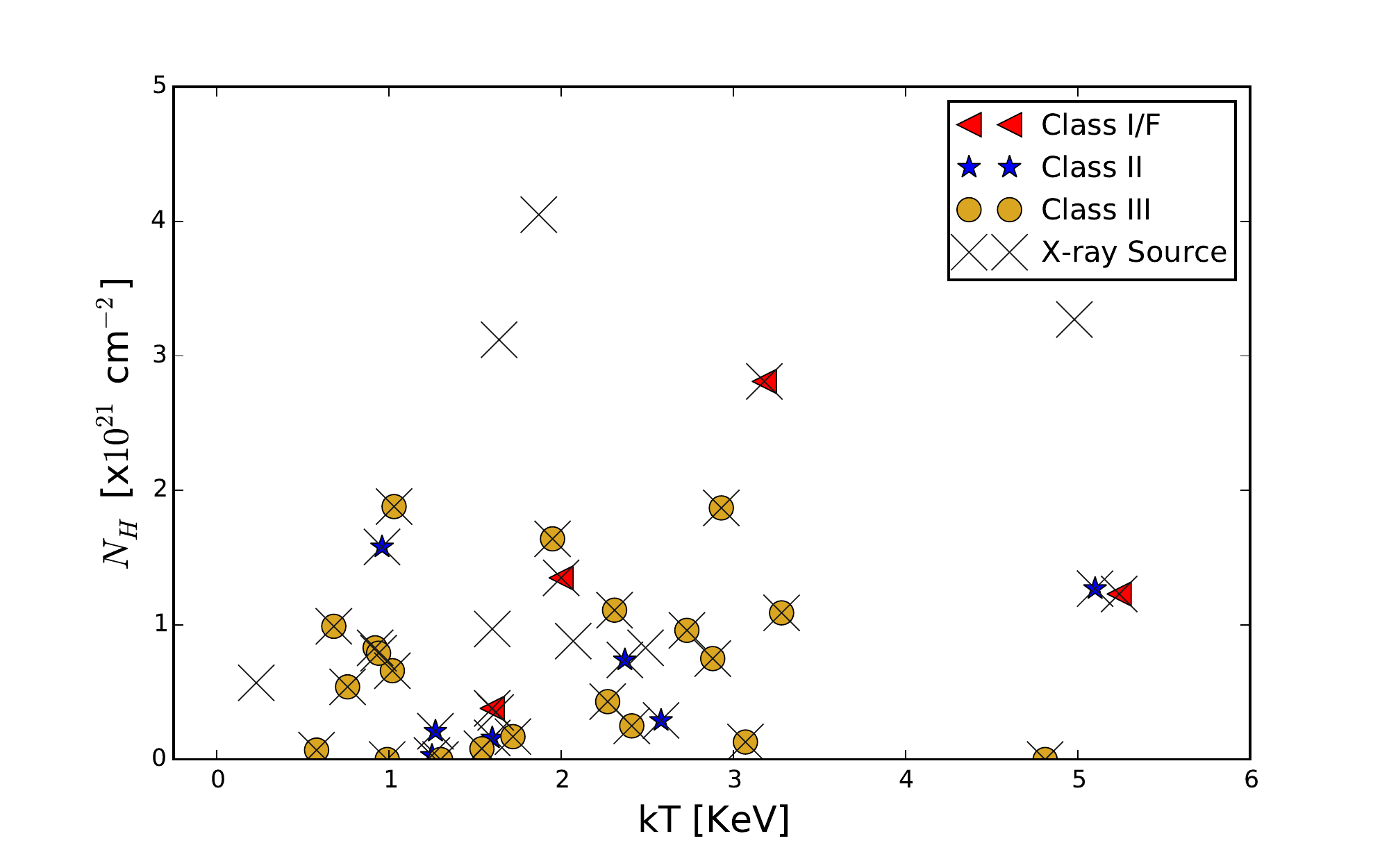}
\caption{
Temperature vs.\ absorption column for X-ray sources that were bright enough for imaging spectroscopy; the legend indicates the IR class. The patterns seen here are common in X-ray studies of regions of star formation. First, Class I sources are faint. Hence, only a few values could be measured. There are several bright X-ray sources which are not identified with a specific IR class; most of these are fairly absorbed.
}
\label{X-ray_Char}
\end{center}
\end{figure*}

Spectral fits were attempted for all X-ray sources with more than 30 net counts. The observed data were fitted to an optically thin collision--less  plasma.  Two fitting algorithms were employed. We typically used the CIAO statistic ``$\chi^2$-data variance" on binned data,
but on sources with net counts between 30 and 50 the resultant $\chi^2$  fits were very poor and the C-Statistic \citep{Cas79}, designed for unbinned data gave more reliable results.  While there were 47 X-ray sources with over 30 counts and IR matches, only 27 spectral fits were considered `good', in the sense that the $\chi^2$/degree of freedom ($dof$) was between 0.5 and 1.5. In general, the sources were found to be lightly absorbed with a fitted value of \nh $< 10^{21}$cm$^{-2}$ for more than half of 
the sources and only one source exceeding \nh= 5.0 $\times 10^{21}$cm$^{-2}$ (Figure~\ref{X-ray_Char}).  The resultant fitted $N_H$ values of the ``good'' were well behaved in the sense that the mean (0.97 $\times 10^{21}$cm$^{-2}$) was similar to the  median (0.79  $\times 10^{21}$cm$^{-2}$) and the standard deviation (0.90  $\times 10^{21}$cm$^{-2}$) was similar to the median absolution deviation  (MAD; 0.67 $\times 10^{21}$cm$^{-2}$). 

Temperatures had a wider range in values since $Chandra$ is not very sensitive to temperature  differences  when the temperature is above 10 keV (This means that the 
 $\chi^2$/$dof$ for a temperature of 10 keV is not too different from 25 or 50 keV).  So while most of the measured temperatures were below 2 keV, a few best fit temperatures exceeded 10 keV, but none of those hottest sources were associated with $Spitzer$ objects. The median values of kT=2.1~keV, \nh=0.8$\times 10^{21}$ cm$^{-2}$ are the more appropriate metrics to give a sense of the typical values.  As such these temperatures are similar to other low mass stars in fairly young clusters \citep[e.g., RCW~38, IC 1396;][]{Wol06,Get12}.  This stands somewhat in contrast to the $N_H$ values which are fairly low for such a young cluster - i.e., few highly absorbed sources. 

X-ray selected samples in star forming regions provide samples of cluster members, unbiased with respect to their IR properties. But they are not unbiased altogether, since an object's characteristic X-ray luminosity is typically a nearly fixed fraction of the bolometric luminosity; X-rays only sample the brightest part of the photometric sample.  Indeed, well over 90\% of the X-ray sources in the overlap field are brighter than [3.6]$<$14.  Following \citet{Sie00}, this is fairly close to the stellar/brown dwarf boundary for unabsorbed YSOs at a distance modulus of about 10.  
From the same models, such a star is about the diameter of the Sun but with a little less than 0.1 $L_\odot$.  The $Chandra$ observation would have been able to detect such a source assuming log $L_x/L_{bol} > -3.9$.   Furthermore it appears that the initial mass function turns over near the brown dwarf limit \citep{Bas10},  so [3.6]$<$14 sets a fairly strict physical limit on the detection of stellar X-rays from PMS stars associated with GGD~12-15. In the X-ray data, there is then a gap of nearly 2 magnitudes before two additional sources with both X-rays and photometric data are found.  The faintest X-ray source associated with a YSO had a [3.6] magnitude of 14.1,
Two other X-ray sources with IR counterparts at  [3.6]$\sim$16 have unknown membership status, but we suspect they are most likely to be AGN.

\section{Analysis}
\label{sec:Analysis}

There were 1017 objects monitored with $Spitzer$. About 1/3 of these objects are in the overlap field.  The vast majority of these are not YSOs but just field stars or other background objects along the line of sight. As discussed in \S2.2,  to account for the differences among the YSOVAR fields we select a ``standard set'' of (probable) members following Paper~I. In our case the standard set is composed of 102 objects identified as having IR excesses, combined with 39 X-ray sources that have photospheric SEDs, creating a standard set of 141 stars. We also define an ``augmented'' set of probable YSOs which includes the 14 X-ray sources that are brighter than 14.1 magnitude in both [3.6] and [4.5] and which have IR colors consistent with Class~I and Class~II objects.  This should exclude background AGN and gives us 155 stars in the augmented set.

Variable stars are defined in Paper~I as sources with Stetson index $>$ 0.9 or by $\chi^2 > 5$ in either [3.6] or [4.5] lightcurves. We also considered in this group weakly variable periodic sources (see \S~\ref{sec:IRVar}) that do not reach either the $S$ or $\chi^2$ criteria, which represent about 25\% of all the periodic sources and include 4 of the 18 variable Class III objects.

\subsection{Periodic Sources}
\label{sec:per}

We identify 39 periodic sources following the methods described in \S~\ref{sec:IRVar} which leaned on the Lomb-Scargle \citep{Sca82} test with confirmation from several methods to ensure robustness. The periodic sources are listed in Table~\ref{TAB:per}.  In addition to the star name and the period, we include the false alarm probability (FAP) and  the power of the periodogram.  Some periods are much more clear and less affected by linear trends and cycle to cycle variations.  The formal errors on the period determination with linear frequencies is very small when there are 79 samples spaced over 38 days -- about 0.0007 (in frequency space) for a 3$\sigma$ signal \citep[e.g., eq.\ 14][] {Hor86}.  Hence, the precision is a function of both the period and amplitude of each star.  But assuming a 3$\sigma$ signal, typical periods should be precise to better than 0.001 for periods below a day, 0.01 for periods below four days, 0.04 for periods below 7 days and periods of 14.5 days should have precision of about 0.15 days. One caveat is that some of the periods are more stable than others.  This appears in the periodogram analysis as either a wider peak or a higher FAP.  The quantitative meaning of this, in terms a systematic error, is not clear. However, sources in Table~\ref{TAB:per} with higher FAPs have a correspondingly greater uncertainty in their periods.  
Further, the periods are measured by tracing ephemeral phenomena (e.g., spots, disk warps, etc.) so  phase shifts are common among YSOs and limit the ultimate precision of such period determination.  In Table~\ref{TAB:per}, we list the periods to two decimal places with the exception of the one period which is less than 1 day and appears to be an eclipsing binary.

Figures~\ref{Fig:57149} and \ref{Fig:57807} show some of the cleanest periodic lightcurves,  while Figures~\ref{Fig:57065}, \ref{Fig:57136} and \ref{fig:57186} show increasingly complicated periodic light curves.  In Table~\ref{TAB:per}, we also include the SED Class as determined by the available colors \citep{Gut09} and the 2-24 \micron\ slope for comparison. From this we can see that 4 periodic sources had insufficient data for color classification; this acts as a reminder that every source gets an SED slope classification, but only 65\% in this sample have the same classification in both systems. This is consistent with Paper~I which systematically compared results between color and SED classes and found 75\% of the objects receiving the same classification for classified objects. The slope classification, which we use throughout the analysis, cannot distinguish transition objects from Class~II when using the  2-24 \micron\ slope. Two of the periodic YSOs are identified as transition objects by \citet{Gut09}.  
We also track which periodic sources are X-ray sources and which are part of the standard set of cluster members based on the definition in \S2. 

\skipthis{

error = 3sigma/ 4N^.5TA
A~3sigma 
1/4*79^.5*40 days 
= 0.0007 days ^-1

for orion 
150^.5 *800

\begin{deluxetable}{lccr}
\tabletypesize{\small}
\tablecaption{\label{Tab:per} Properties of periodic sources.}
\tablehead{\colhead{ID} & \colhead{Name} & \colhead{Unit}  & \colhead{Comment}}
\startdata
1 & RA & deg &  J2000.0 Right ascension \\
2 & DEC & deg &  J2000.0 Declination \\
3 & IAU\_NAME & J2000.0 &   IAU designation  \\
4 & Period & days &  Period \\
5 & Max Power & -- &  LS Power of the most significant peak\\
6 & FAP  & -- &  False alarm probability\\
7 & IRclass & -- &  IR class according to \citet{Gut09} \\
8 & SEDclass & -- & IR class according to SED slope  \\
9 & cmd\_alpha\_36\_45 & rad & angle of best-fit line in CMD \\
10 & cmd\_alpha\_error\_36\_45 & rad &uncertainty on angle \\
11 & X-ray & -- & $Chandra$ counterpart ? \\
\enddata
\tablecomments{This table is published in its entirety in the electronic version of the journal. Here the table columns are described as a guide to form and content.}
\end{deluxetable}
}

\begin{deluxetable}{lccrrrcc}

\tabletypesize{\footnotesize}
  
  \tablecaption{Properties of Periodic Sources.\label{TAB:per}}
  \tablewidth{0pt}

  \tablehead{
    \colhead{IAU Name} &
    \colhead{X-Ray} &
    \colhead{Standard Set} &
    \colhead{Period} &
    \colhead{Peak} &
    \colhead{False }&
        \colhead{Class from}  &
    \colhead{Class from} \\
     \colhead{~} &
    \colhead{Source} &
    \colhead{of Cluster} &
    \colhead{~} &
    \colhead{~ } &
    \colhead{Alarm }&
    \colhead{Gutermuth et al.} &
        \colhead{SED} \\
           \colhead{SSTYSV ...} &
    \colhead{~} &
    \colhead{Members } &
    \colhead{[days]} &
    \colhead{[power] } &
    \colhead{Probability }&
    \colhead{(2009)} &
        \colhead{Slope} 
   }
\startdata
  J061034.11-061711.8 & no & yes & 5.80 & 18.0 & $<0.001$ & II & II  \\
  J061034.87-061316.8 & yes & yes & 5.88 & 24.2 & $<0.001$ & III & III \\
  J061035.46-062024.9 & no & no & 4.15 & 13.6 & 0.009 & \ldots & III \\
  J061035.92-061249.8 & no & yes & 2.52 & 13.4 & 0.012 & Transition & II \\
  J061036.97-061158.6 & no & yes & 7.71 & 13.0 & 0.019 & Transition & II \\
  J061037.00-061317.6 & no & yes & 12.83 & 14.7 & 0.003 & II & III \\
  J061039.39-061717.3 & yes & yes & 5.47 & 16.3 & 0.001 & III & III \\
  J061039.80-061310.8 & no & yes & 10.32 & 15.4 & 0.002 & II & III \\
  J061041.99-061042.5 & yes & yes & 1.34 & 32.9 & $<0.001$ & III & III \\
  J061043.11-061220.6 & yes & yes & 4.64 & 30.6 & $<0.001$ & III & III \\
  J061043.55-061240.6 & no & yes & 5.88 & 12.8 & 0.022 & II & I \\
  J061043.83-061235.6 & no & no & 2.76 & 15.1 & 0.002 & III & II \\
  J061044.25-061110.2 & yes & yes & 5.40 & 30.7 & $<0.001$ & III & III \\
  J061044.80-061202.1 & no & yes & 7.71 & 12.8 & 0.021 & II & II \\
  J061045.86-061135.4 & yes & yes & 7.71 & 13.8 & 0.008 & II & II \\
  J061046.21-061125.0 & no & yes & 4.24 & 13.4 & 0.012 & II & II \\
  J061048.84-061143.8 & no & yes & 6.15 & 14.8 & 0.003 & II & II \\
  J061048.96-061149.8 & yes & yes & 8.92 & 13.9 & 0.008 & II & I \\
  J061049.21-061130.0 & yes & yes & 6.67 & 35.4 & $<0.001$ & III & III \\
  J061049.45-061218.3 & no & no & 4.43 & 13.6 & 0.009 & \ldots & F \\
  J061050.68-061155.5 & yes & no & 13.03 & 12.8 & 0.021 & \ldots & F \\
  J061051.50-061404.1 & yes & yes & 7.71 & 16.9 & $<0.001$ & II & II \\
  J061052.26-061059.6 & no & yes & 6.56 & 14.0 & 0.006 & II & F \\
  J061052.48-061001.5 & no & yes & 10.32 & 20.0 & $<0.001$ & II & II \\
  J061053.12-060935.6 & yes & yes & 13.94 & 14.3 & 0.005 & III & III \\
  J061053.32-061052.0 & yes & yes & 14.00 & 17.5 & $<0.001$ & II & II \\
  J061053.35-061115.3 & no & yes & 12.22 & 12.5 & 0.029 & II & F \\
  J061053.88-061130.7 & yes & yes & 3.11 & 35.1 & $<0.001$ & III & III \\
  J061054.04-061021.0 & no & yes & 10.88 & 13.1 & 0.016 & II & II \\
  J061054.22-061220.1 & no & yes & 5.25 & 22.4 & $<0.001$ & II & II \\
  J061054.86-061107.7 & yes & yes & 4.15 & 15.6 & 0.001 & II & II \\
  J061056.75-061101.7 & yes & yes & 9.34 & 27.3 & $<0.001$ & II & F \\
  J061101.20-061107.0 & yes & yes & 1.08 & 20.8 & $<0.001$ & III & III \\
  J061101.21-060934.6 & yes & yes & 2.40 & 21.7 & $<0.001$ & II & II \\
  J061103.23-061016.8 & yes & yes & 5.71 & 22.9 & $<0.001$ & II & II \\
  J061104.66-060057.9 & no & no & 0.146 & 34.6 & $<0.001$ & \ldots & III \\
  J061105.20-061156.2 & yes & yes & 0.64 & 15.6 & 0.001 & III & III \\
  J061107.12-060807.7 & yes & yes & 3.52 & 25.0 & $<0.001$ & III & III \\
  J061110.61-060606.2 & no & no & 4.97 & 12.4 & 0.016 & III & III \\
 \enddata
\end{deluxetable}

\begin{figure*}[htbp]
\begin{center}
\includegraphics[width=6.5in]{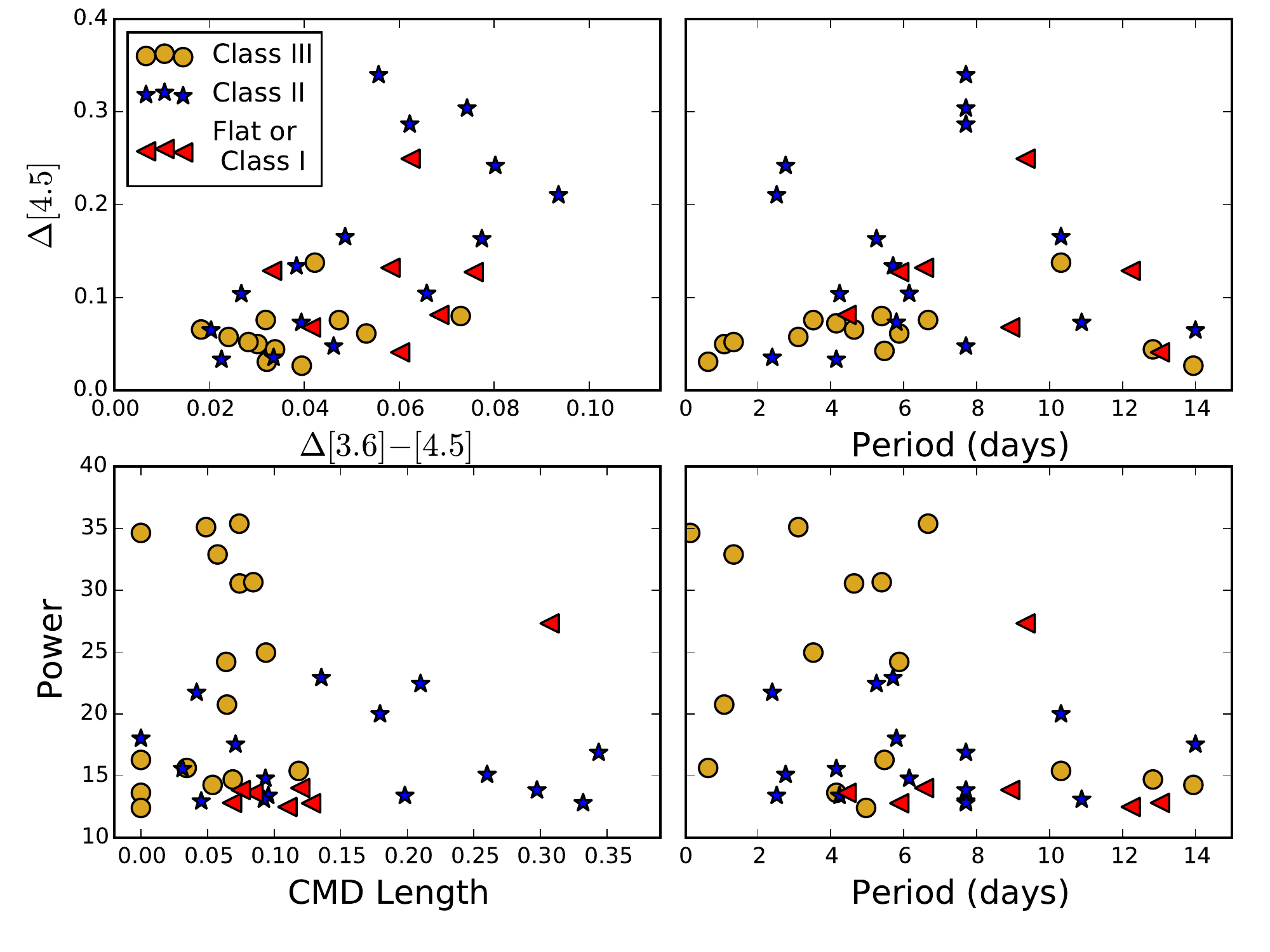}

\caption{ 
Scatter plots of the key IR parameters of periodic sources.  In all plots,  blue stars indicate Class II objects, gold circles indicate X-ray sources with SED slopes appropriate for photospheric SEDs and the red triangles are objects with flat or Class~I SEDs.
It appears in the upper plots that Class II objects have, on average,  larger changes in color and magnitude
compared to Class III objects. The Class I/F objects are just a few, and show comparable color and magnitude variability as Class II.
The right hand plots show that diskless objects have, on average, shorter periods with more inherent power. 
}
\label{fig:scatter:per}
\end{center}
\end{figure*}

\begin{figure*}[htbp]
\begin{center}
\includegraphics[width=6.5in]{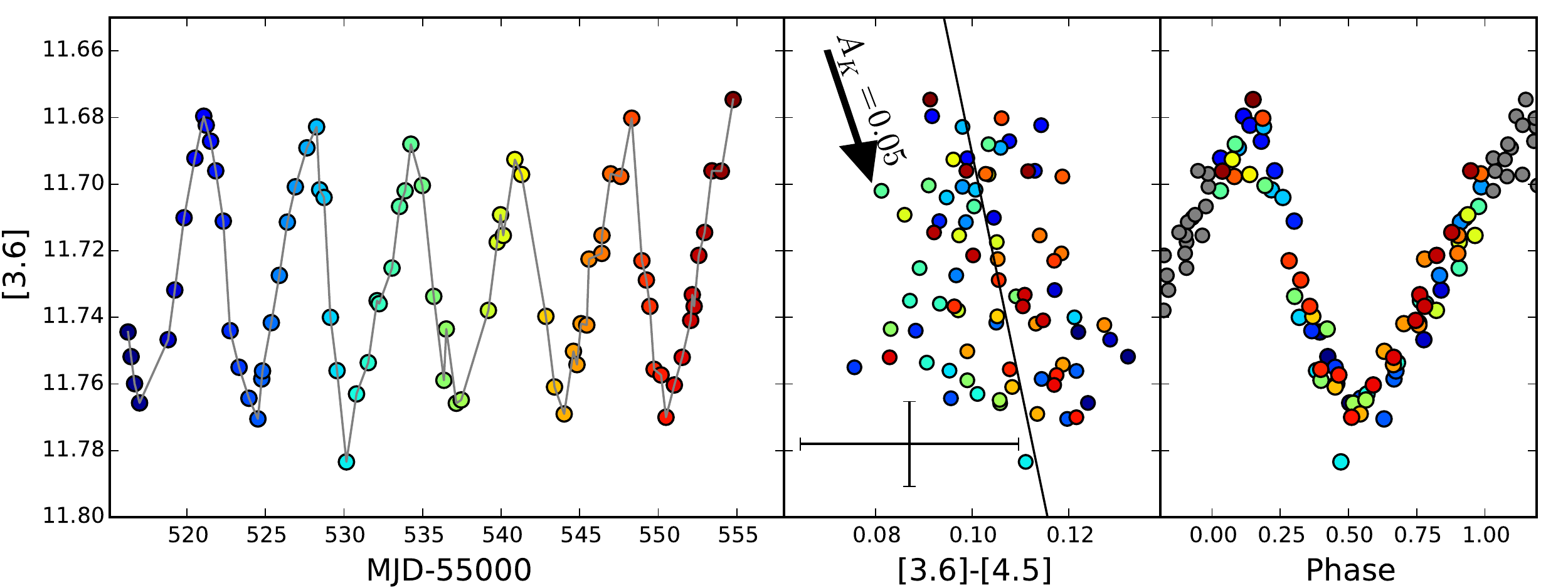}
\caption{
Lightcurves for the Class~III object SSTYSV J061049.21-061130.0. In the left panel, we show the raw lightcurve for [3.6].  Time is in units of Modified Julian Day $-$ 55000. In the center panel, we plot the [3.6]-[4.5] color against the [3.6] magnitude for the source, The solid line is a fit to the data with a slope of about 82$\pm3^{\rm {o}}$ The arrow indicates an $A_K$=0.05 reddening vector, adapted from \citet{Ind05} and the typical errors are indicated near the bottom.  The folded period is shown in the right panel. The colors of the dots change with time, blue at early time going to red at late time. In the phased panel, more than one full phase is shown for clarity, the repetitive data are shown as grey.}
\label{Fig:57149}
\end{center}
\end{figure*}

\begin{figure*}[htbp]
\begin{center}
\includegraphics[width=6.5in]{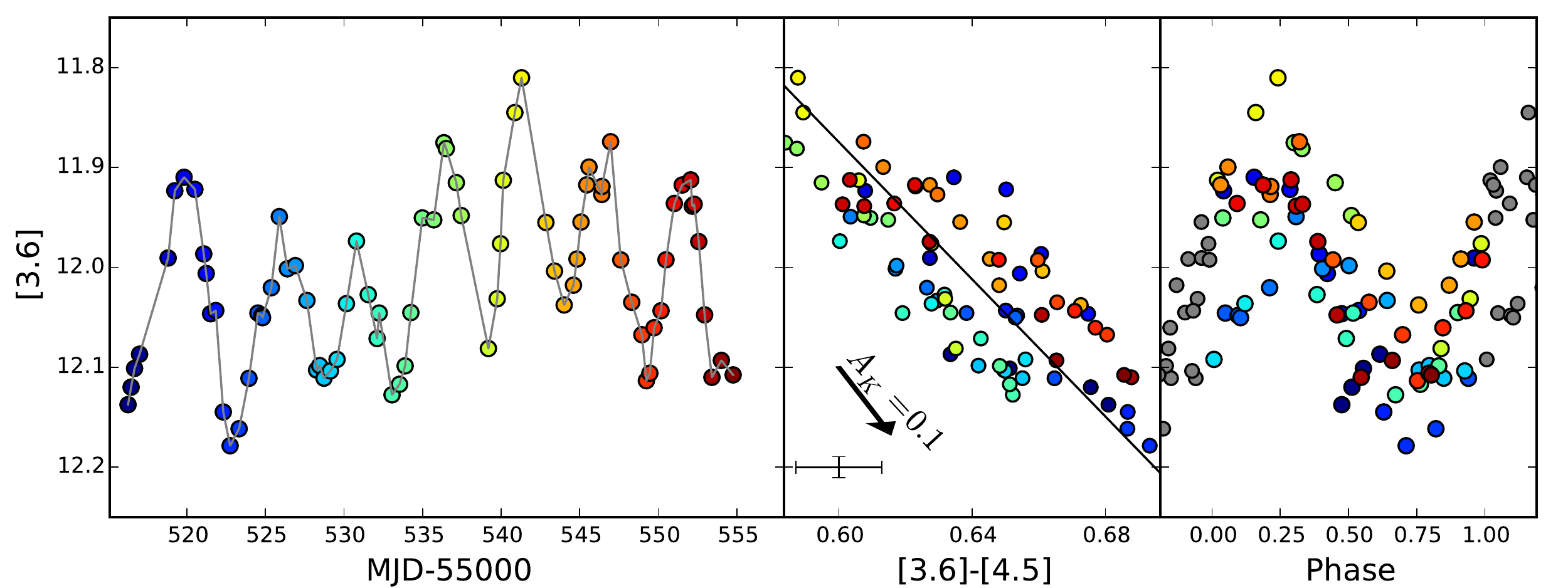}

\caption{
Lightcurves for the Class~II object SSTYSV J061054.22-061220.1.  Symbols and panel descriptions are the same as the previous figure. The folded data do not overlap exactly. The color variances are consistent with changes in reddening. The fitted line has an angle of about 74$\pm2^{\rm {o}}$,   which is parallel to the reddening vector  that has also slope of about 74$^{\rm {o}}$ and a length (in this case) of $A_K$ = 0.1.}
\label{Fig:57065}
\end{center}
\end{figure*}

\begin{figure*}[htbp]
\begin{center}
\includegraphics[width=6.5in]{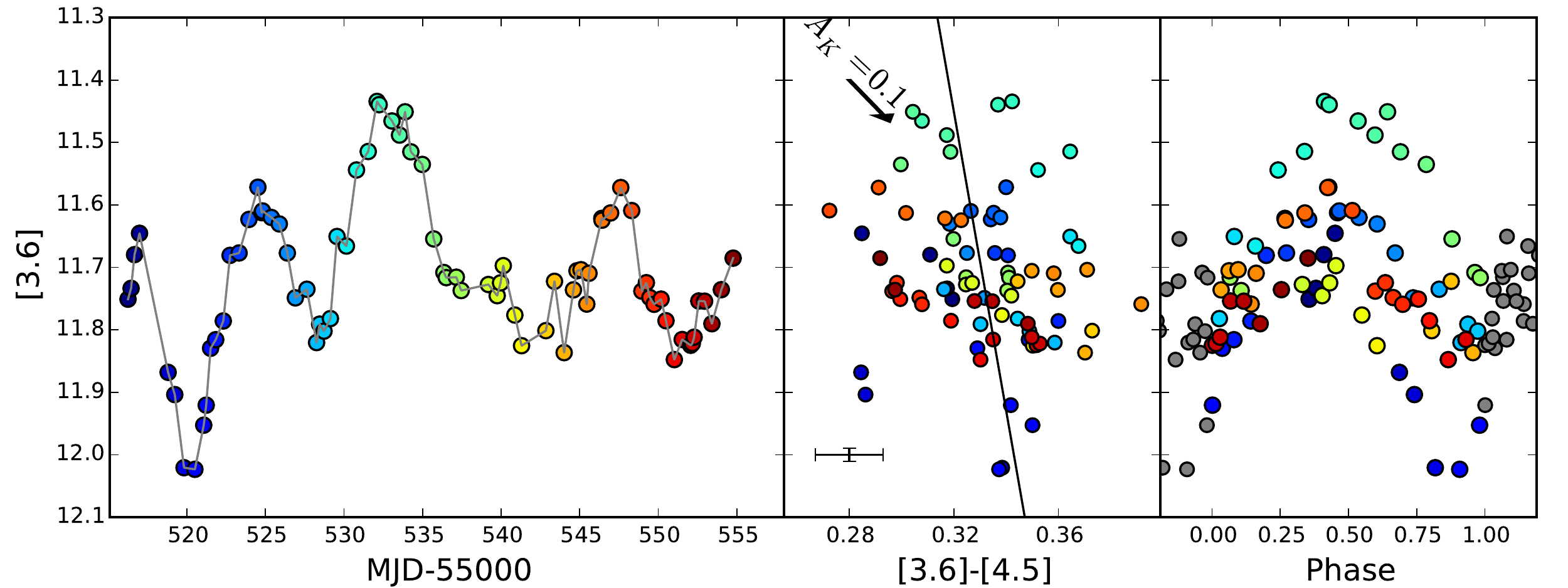}

\caption{
Lightcurves for the Class~II object SSTYSV J061045.86-061135.4 which has a complex lightcurve. Symbols and panel descriptions are the same as the previous figures.
The fitted line is about 88$\pm2^{\rm {o}}$.
 The reddening vector is $A_K$ = 0.1.}
\label{Fig:57136}

\end{center}
\end{figure*}

\begin{figure*}[htbp]
\begin{center}
\includegraphics[width=6.5in]{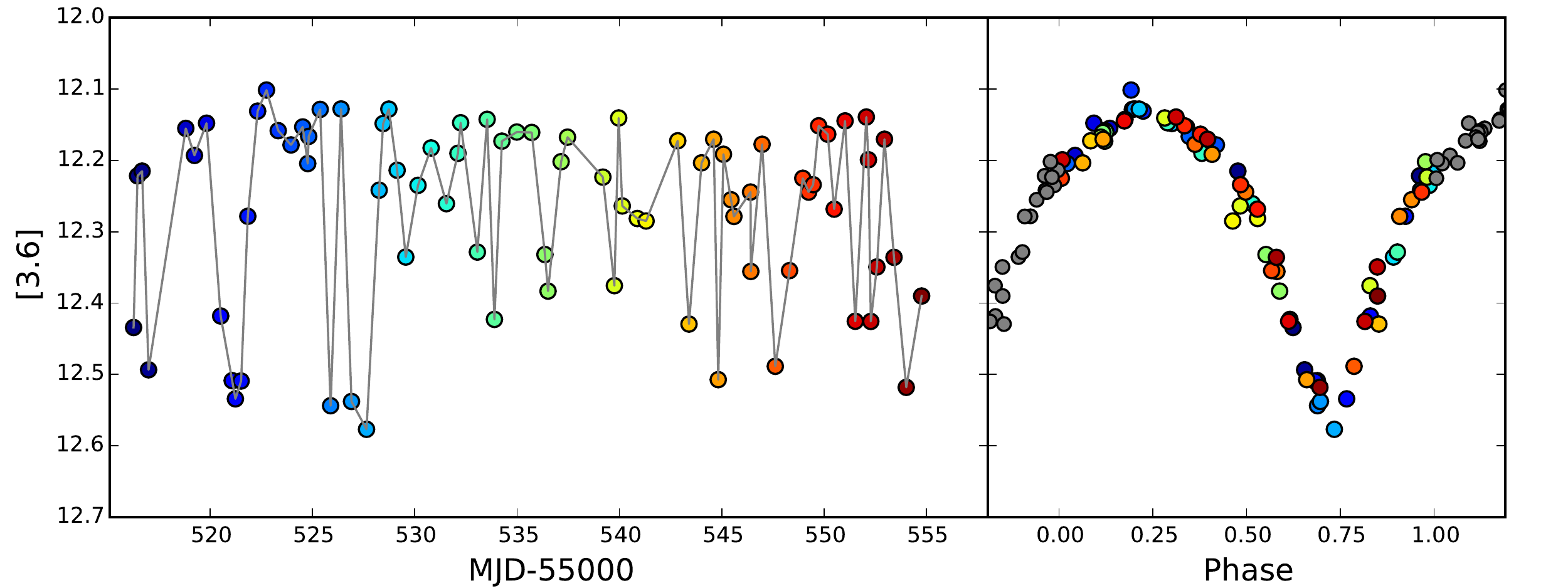}

\caption{
Raw (left) and folded (right) 3.6\micron\ lightcurves for SSTYSV J061104.66-060057.9. Colors are derived from the date of the observation as in the previous figures.   The short (3.5 hour) period,  rounded top and sharp bottom are typical of eclipsing systems. }
\label{Fig:57807}

\end{center}
\end{figure*}

We divided the periodic sources into 3 groups, derived from IR Class: those with flat or Class I SEDs, those identified as Class II,  and those with no excesses relative to a photospheric SED -- Class III. This latter category may include non-YSOs in the field which happen to be strongly periodic.  A probable example of this  is SSTYSV J061104.66-060057.9 which appears to be a contact binary system with an orbital period of less than three and a half hours. SSTYSV J061104.66-060057.9  is neither an X-ray source, nor in the standard set of cluster members.  We compared many traits among the Classes,  but focused on the distributions of the periods, the color and magnitude ranges and the quality of the periodic signature.  We used both scatter plots for this comparison (see Figure~\ref{fig:scatter:per}) and two--sided Anderson-Darling tests, which would tell us if any trends we perceived by eye were statistically significant.

There are several trends which are present at a probability exceeding 90\%.   
First, the measured period also has a weak dependence on Class, with the Class II objects having, on average, the longer periods. 
Second, the observed brightness change is different at the 2$\sigma$ level with the Class II objects generally having a larger change than Class III. There are several metrics that could be a proxy for brightness change. We find the relation between periodogram power and overall change is strongest when calculated in terms of the length of a line-segment fitted to the CMD which includes the central 80\% of the data (\ie 10\% outliers are removed; see \S~\ref{sec:noper}).  While measuring the length of the line-segment fitted to the CMD  maximizes the amount of change measured, the effect that  Class II objects change more than Class III objects also was present in other proxies.   The most significant trend is the distributions of the periodogram power with Class. Periodogram power can be treated as a proxy for the stability of the periodic signal over the 38 day observation window,   Class III objects have more periodogram power (in the channel identified with the most probable period) than other classes of objects.  In other words, Class III objects have the most unambiguous periods.  

The strongest trends were found when we combined all the disked stars (Flat Spectrum, Class I and Class II) and compared them with the disk-free population (Class~III). Among the quantities shown in Figure~\ref{fig:scatter:per} we found the magnitude of the change differed at a confidence exceeding 99.9\%.  The distribution of the peak power, which is a reflection of both the stability and the shape of the lightcurves differed at about 99.5\% confidence.

It has been long known that diskless YSOs have more rapid rotation periods, on average, than Class II sources \citep{Bou86}.  This has been linked to the interaction of the magnetic field of the star with the ionized inner disk slowing the stellar rotation \citep[][and many others since]{Edw93}.  The difference among the variations (i.e., $\Delta$) is generally attributed to the fact that Class III sources only have a sole mechanism of variation -- dark spots -- and the global impact of dark spots is expected to be limited on theoretical grounds to no more than 15\% and empirically about half that \citep{Car01, Wol13, Sch05}.  Class II objects, on the other hand, are hybrid systems that include a disk, accretion, as well as a star with both bright and dark spots. In principle, the disk alone can block all the light from the star.  This also explains why the diskless sources have less scatter from the periodic signal as manifest by the higher power in the periodic signal.  Given this, it is a little surprising that the observed color changes {\em do not} have a stronger trend -- however the sample size is very small. The observed color changes in the Class II periodic sources are small ($<$9\%) when compared to color changes seen in $H-K$ for periodic Class II objects \citep{Wol13} and the GGD~12-15 [3.6]$-$[4.5] variable sample as a whole (see \S~\ref{sec:noper}).

In the following,  we discuss a few sources which act as examples of recurrent observed behavior.   Many of the Class III periodic sources are fairly similar. They have very clear periods, with little deviation beyond the known noise,  and periods of a little under a week. Figure~\ref{Fig:57149} shows an example Class III source (SSTYSV J061049.21-061130.0).  Notice that the color variation  [3.6]-[4.5] is fairly small, $\le$ 0.05 mag.  
Following \citet{Gun14} we fitted the distribution of points in color-magnitude space using a linear regression which included the color, [3.6] magnitude and associated errors. The results of the fits are slopes, with errors that are included in Table~\ref{TAB:IR}.  In this case, the fit is nearly vertical and the data points highly scattered indicating little color change with brightness. 
 Over 38 days, more than 4 cycles, the scatter of the phase-folded lightcurve is about 0.02 mag, which is about the observational error. In the CMD,  there is no skew in the color as a function of brightness, although the dispersion is higher at the faintest measurements.     This is one of five YSOs, in GGD~12-15, known to be associated with a centimeter wave radio source \citep[VLA~8;][]{Gom02}.  The radio emission is time variable as well, but has only been measured on much longer timescales.

There is more interesting diversity among the lightcurves of periodic Class II sources. Figure~\ref{Fig:57065} shows the Class II object SSTYSV J061054.22-061220.1. Again the period is very pronounced in the lightcurve.  
The color-magnitude diagram shows the changes in color and magnitude are linearly correlated and with a fitted slope of about 74$\pm2^{\rm {o}}$, nearly the same as the reddening vector \citep{Ind05}. This suggests that the flux changes are caused by dust with a particle size distribution similar to interstellar dust. In the middle panel,  while the data as a group follow the reddening vector,  we note a secondary trend perpendicular to the reddening vector, such that most of the data from days ...525-535 are consistently  bluer and fainter  than corresponding data from the first or last 10 days. This is apparent when  the folded data show that the various cycles do not overlap exactly.  The data from the first $\approx$ 2 cycles are about 0.15 mag below the data from the next couple of cycles with the final cycle in between.  


For many of the Class II objects, the situation is far more complicated.  We use as an example SSTYSV J061045.86-061135.4 (Fig.~\ref{Fig:57136}).    In this case, the 7.71 day period found by the computer is obvious to the eye in the form of 3 sharp peaks at days ...524, ...532 and ...547 respectively.   The expected  peak near day ...540 is barely visible in the [4.5] channel above a general downward trend.  The fitted slope in color-magnitude space is nearly vertical, 89.8$\pm2^{\rm {o}}$. Here we note from the CMD that the overall lightcurve was faintest in the first and last week and this seems to drive the fit.  In the last weeks (orange and red in the CMD plot) the data seem to fit on an ISM-like reddening curve.   The same can be said of the data from the middle $\approx$ 2  weeks except that the overall pattern is shifted about 0.2 mag brighter and  0.03 mag redder.  Similar effects have been seen in the NIR \citep{Wol13}.  The cycle-to-cycle variations reach 0.3 mag.
 
SSTYSV J061104.66-060057.9 appears to be a contact binary system (Fig.~\ref{Fig:57807}).  We assert this based on the extremely short period of  less than 3.5 hours, which would exceed breakup velocity for a rotating sub-giant star. Indeed there is no evidence it is a subgiant PMS star, its colors are consistent with it being a diskless mid-K star which has no extinction, and no X-rays were detected.  On the other hand, the star has $\Delta 3.6$ of about 0.5 mag, very low scatter and an asymmetric lightcurve with a sharp bottom and relatively rounded top.  SSTYSV J061104.66-060057.9 is also located north of the main body of GGD~12-15 and hence only detected in the [3.6] channel. The other rapid rotator is SSTYSV J061047.75-061200.9, a probable Class III which has a 15.3 hour period.  While fast, this is no faster than previously known ultra fast rotating YSOs (\eg\ Bouvier \e 1993).   SSTYSV J061047.75-061200.9 has $\Delta$ of about 0.03 mag with a slight reddening, but no noise on the signal above the expected systematic noise.  In fact, the sinusoidal signal in this case is so weak that, had it been complicated by a disk signature, the periodicity would have been difficult to identify. 

\subsection{All Variable Sources}
\label{sec:noper}

   
The total number of variables according to all criteria is 148.  This includes 106 of the 141 stars from the standard set of cluster members.  Put another way, 75$\pm$4\% of the standard set of members are variable.\footnote{In this section, errors are established using a $\sim 68\%$ binomial confidence interval \citep{Wil27}. The asymmetry in the confidence interval is averaged out for simplicity.}  The standard set includes 69 X-ray sources, of which 47 (68 $\pm$ 6\%) are variable.   Amongst the flat SED and Class~I sources,  33 of 36 (92 $\pm$ 4\%) are variable as are 55 of 66 (83 $\pm$ 5\%) objects with Class II SEDs.
Just under half of the X-ray sources with Class~III SEDs are variable as well  (Table~\ref{TAB:STD}).  

\begin{figure*}[htbp]
\begin{center}
\includegraphics[width=6.0in]{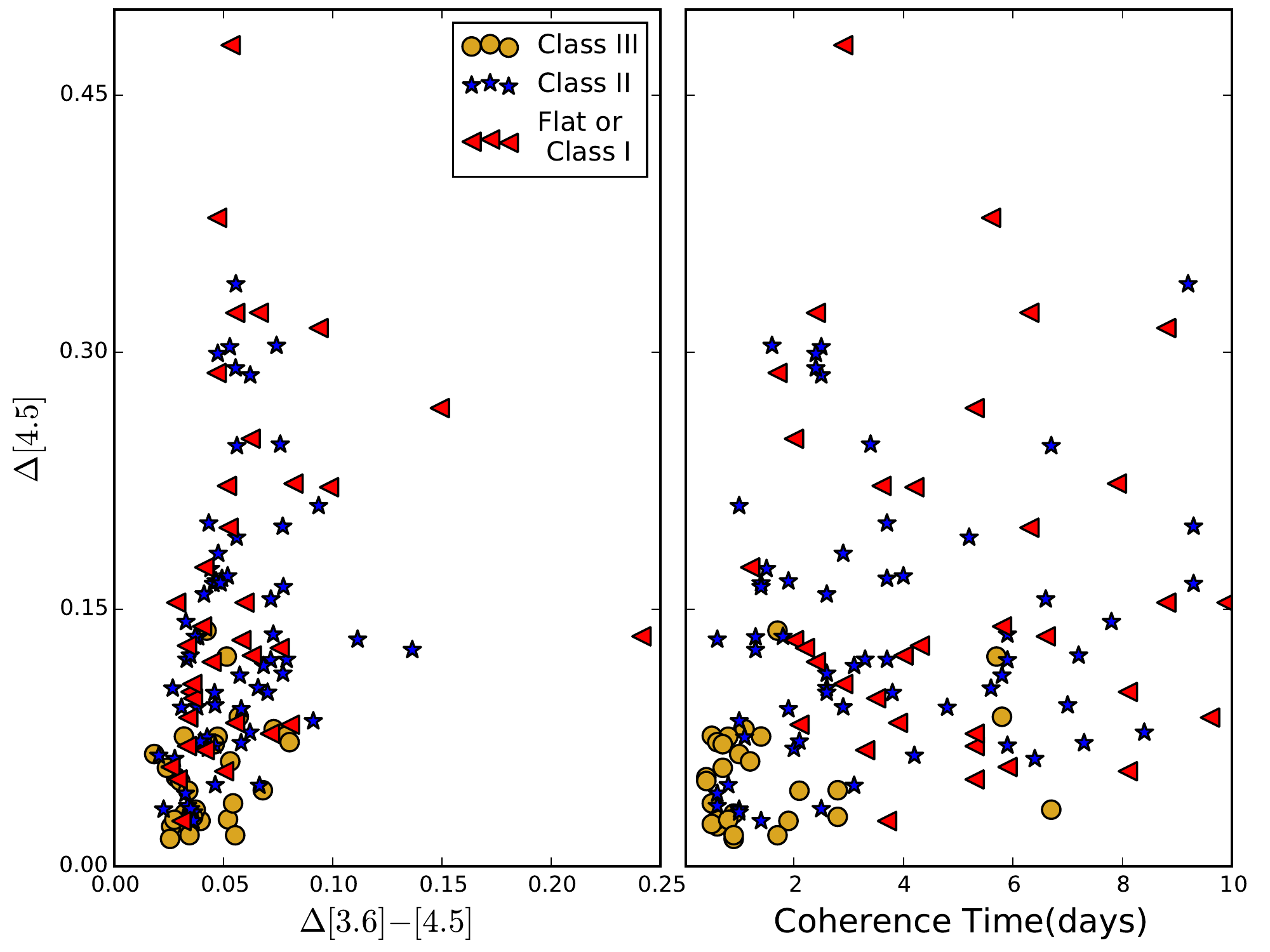}
\caption{ 
Scatter plots of the key IR parameters for variable sources.  In all plots,  blue stars indicate Class II objects, gold circles indicate X-ray sources with SED slopes appropriate for photospheric SEDs and the red triangles are objects with flat spectrum or Class~I SEDs.
The Y-axis in both plots is the range of change of values at [4.5] with the upper and lower 10\% of the data points removed.
The X-axis of the left hand plot is the change in the [3.6]-[4.5] values with the same outliers removed.  Very few sources show extreme color changes, but those that do are as likely to be Class I/F as Class II despite the fact that there are more Class II sources in the sample. 
The X-axis in the right hand plot is ``coherence time" (i.e., the value of $\tau$ when the ACF($\tau$) = 0.25).  It is clear that the Class III sources have the shortest coherence time.  The longest coherence times are possessed by Class I/F sources.
}
\label{fig:scatter:var}
\end{center}
\end{figure*}

In analogy with Figure~\ref{fig:scatter:per}, we examined the distribution of color change and coherence time as a function of magnitude change for the various classes.  Figure~\ref{fig:scatter:var} shows clearly that the Class~III variables show the smallest change.  The other classes show a lot of overlap, but the Class~I and flat spectrum sources seem the most likely to have large, slow changes (high values of coherence time).  Unlike the solely periodic sample, we find a large number of highly variable disked objects with small color changes. The overall trend of larger variability being correlated with larger color change persists. 
The Class I and flat spectrum objects show the largest variations. 
Class~III sources have the shortest coherence time, meaning that they change the most rapidly. We note that many sources have coherence time in excess of 4 days.  In other words, they change, but relatively slowly.  Following the reasoning in the previous section -- wherein we found coherence time was about one-fifth of the period -- {\em if } these stars are periodic and have coherence time $> \sim$3 days, the observation window was too short to identify the periodic nature.

We checked to see if X-ray objects were preferentially variable and found, among the Class I, Flat, and Class II X-ray sources, 29/30  {\em were} variable.  While the rate seems impressively high, sampling errors admit a variability fraction as low as 92\%. Still, this is slightly higher than the overall variability rate of the Class I/F/II sources overall which is consistent with being as high as 90\%.   A two-sided Anderson-Darling test does not indicate that X-ray sources are significantly more variable in the IR. Broken down by class, it is clear that the Class~I sources are the most likely to be found to be variable.  Further, Class~I sources also show the largest change in magnitude.  Flat spectrum sources show the second highest variability fraction, but the median change seen in each channel is smaller than the median changes seen for the Class~IIs.  However the {\em average} change seen in each channel is larger than the average changes seen for the Class~IIs indicating at least some flat spectrum objects are still highly variable. 

In Table~\ref{TAB:STD}, we evaluate the basic metrics of variability.  
The first column lists the subset of interest.  The second column indicates the number of sources of each Class. The third column lists the number of variables as determined by either by periodicity, the reduced $\chi^2$ test or the Stetson test. The remaining columns list statistics of the variable sources in the field using the values of $\Delta$ [Channel]. This is the width of the distribution of photometric measurements for each source from 10\% to 90\%, thus it excludes outliers including flares and eclipses. The final column indicates the median/average Stetson index. The ``No Disk'' row notation here refers to standard set members which are X-ray sources which were {\em not found to have evidence for disk excess} by \citet{Gut09}.  These sources tend to show less variability ($\Delta$) and much lower Stetson indices than the full standard set of Class III objects which, as defined, can include sources which have photosphere SEDs {\bf and} color-excesses as determined by  \citet{Gut09}.  
 A second result is that  Class II and I sources are much more often found to be variable and show larger changes than either the No Disk or Class III groupings. This result is similar to what is seen in the young cluster Lynds~1688 \citep{Gun14}.  
 



\begin{deluxetable}{lrrrccc}

  \tabletypesize{\small}
  
  \tablecaption{Variability Metrics for Subsamples\label{TAB:STD}}
  \tablewidth{0pt}

  \tablehead{
    \colhead{Subset} &
    \colhead{Number } &
    \colhead{\# Vars. } &
    \colhead{\%} &
    \colhead{$\Delta 3.6 $[med.]$^a$} &
    \colhead{$\Delta 4.5 $[med.]$^a$}&
    \colhead{$S$[med.]$^a$} \\
     \colhead{~} &
    \colhead{~ } &
    \colhead{~ } &
    \colhead{~} &
    \colhead{[avg.] } &
    \colhead{[avg.] }&
    \colhead{[avg.] }
   }
\startdata
Standard	 & 141&	 106	&75&	0.116 &0.120&3.11\\
set                   &            &        &       &0.145&	0.140&	3.96\\
\hline
Class I          &  17 &  16  &  94	 &  0.154	 &  0.141 &  3.11\\
~                         &            &        &       &0.197&	0.184&	5.34\\
\hline
Class F          &  19	 &  17  &  89	 &  0.121	 &  0.132 &  3.78\\
~                         &            &        &       &0.165&	0.161&	4.48\\
\hline
Class II  & 66 & 55	& 83 & 	0.138 	& 0.134 & 	3.42 \\
~                         &            &        &       &0.144&	0.145& 4.14\\
\hline
Class III	 &39  &18	 & 46 &	0.069 &	0.063  &1.31\\
~                         &            &   &&0.073&	0.065 &1.46\\
\hline
\hline
Augmented  	 &155	 &116	&75  &	0.117 	 &0.120 &2.95\\
set                         &            &        &       &0.146&	0.142&3.82\\
\hline
Augmented  &	24	& 22& 	92	& 0.174 & 	0.140	& 2.77\\
Class I                     &     &  &       &0.194&	0.189&4.61\\
\hline
Augmented  &	20	& 18& 	90	& 0.118 & 	0.130	& 3.75\\
Class F                     &     &  &       &0.159&	0.154&4.23\\
\hline
Augmented	&  72&  58&  81&  0.136 &  0.132&  	3.36\\
Class II           &            &        &       &0.143&	0.144&4.06\\
\hline
\hline
No Disk	&  35&  14	&  40 &  0.061 &  0.059&  	1.26\\
~                  &    &        &      &0.062&	0.057&1.21\\
\hline
\hline
All others  &201	&21	&10&	0.193&0.192 & 2.13\\
with [3.6] \&[4.5] 	&              &       &       &     0.229& 0.226&1.96\\
\hline 
[4.5] only &	319	&7 & 	2	&\dots&	0.478 & \ldots\\
 ~                         &            &        &     &~  & 0.561 & ~\\
 \hline
[3.6] only&	342&	4&	1&	0.202 &\ldots&\ldots\\
 ~                         &            &        &     &0.203&~&~\\		
\enddata
\tablecomments{$^a$These columns list two metrics for the measured quantities ($\Delta$[3.6], $\Delta$[4.5] and Stetson index respectively). The top value is the {\it median} of all the variables of the given subset.  The bottom value is the {\it average} value of all the variables of a given subset. }

\end{deluxetable}


Within the standard set of members, there are only 14 sources with SED slope indicative of warm circumstellar dust not found to be variable. Of those, four were only monitored in 1 Channel and only one of those four had more than 40 observations. The lightcurve of this object, flat spectrum source SSTYSV J061057.83-061439.7, has a reduced $\chi^2$ to a constant source of $\sim 4.5$ and so was just contained outside our definition of strongly variable.  Of the remaining 10, half have $0.6 < S < 0.9$ and so are on the cusp of our variability criteria. In particular, Class II objects SSTYSV J061045.47-061228.4 and SSTYSV J061043.23-061040.7 are very faint with mean [3.6] $\sim 15.0$.  In both cases, this drives up the observational errors and the requirements on variability. Similarly, Class II source SSTYSV J061053.76-061042.8 has a mean [3.6] of 15 and mean [4.5] of 14.1 and so a greater amount of relative variability is required in this case compared with the typical source. This was simulated in detail in Paper~I (\S 5.4).  The simulations showed that variability detection is a function of both signal-to-noise and the number of samples which diverge from the mean with the typical S/N required being about 4 in a non--periodic case and 2 in a periodic case. Meanwhile, SSTYSV J061044.52-061211.9 is a true victim of circumstance with $S=0.86$ which is within rounding error of our 0.9 criterion. It is reasonable to conclude that these 10 sources are consistent with being variable as well.  

\begin{figure*}[htbp]
\begin{center}
\includegraphics[width=6.5in]{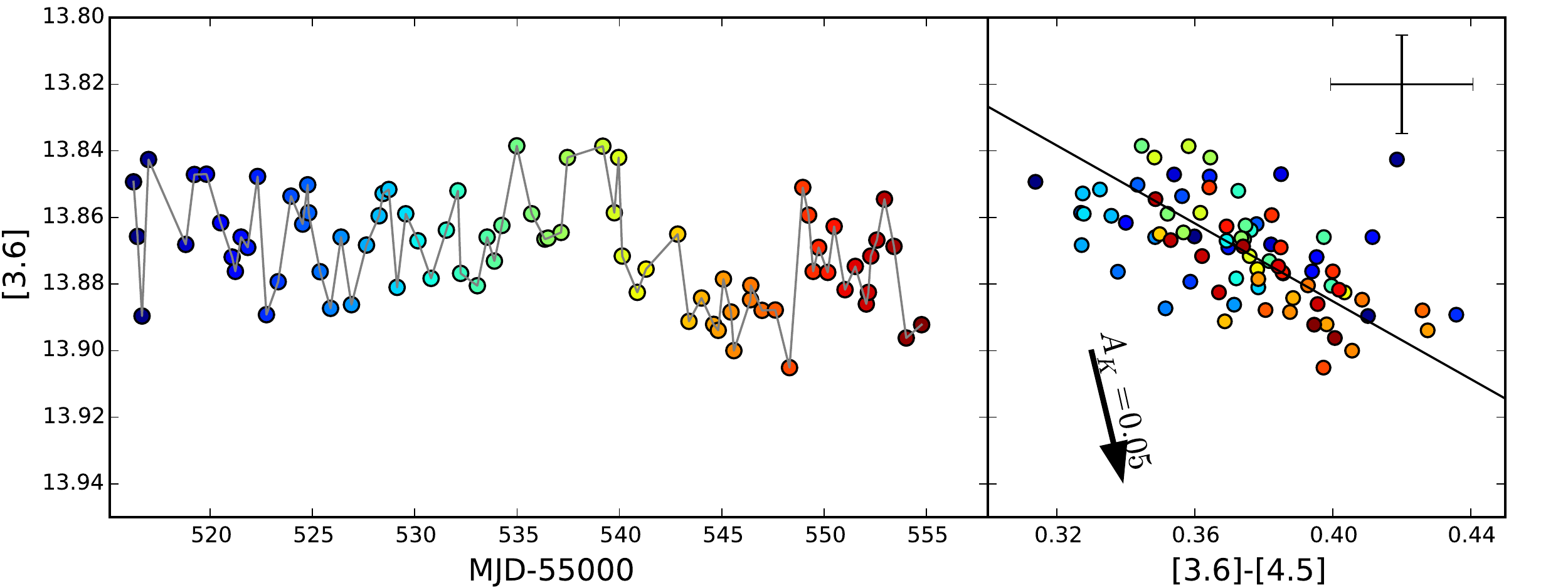}

\caption{
Lightcurve for the non-variable Class~II object SSTYSV J061045.94-061115.7. Symbols and panel descriptions are the same as the previous figures.  The overall change in [3.6] is only slightly greater than the 1$\sigma$ error bar shown in the left-hand figure. 
While the change is correlated between the two channels, the fitted slope, about 30$\pm3^{\rm {o}}$, is much flatter than 
 the reddening vector ($A_K$ = 0.05) or other more variable objects.}
\label{Fig:57168}

\end{center}
\end{figure*}

However, the other three objects, Flat SED source SSTYSV J061049.80-061144.8  and Class II sources SSTYSV J061044.23-061221.9 and SSTYSV J061045.94-061115.7 (Figure~\ref{Fig:57168}) appear to be truly non--variable or minimally variable.  The Flat spectrum source is quite bright (mean [3.6]$\sim 9.5$) and yet the changes seen are less than 5\% ($\Delta$) over the entire period.  This is similar to the overall changes seen in the two non-variable Class II sources. 

\subsection{Beyond the Standard Set of Cluster Members}
\label{sec:aug}

In addition to the standard set of cluster members, we identified 14 sources as probable YSO based on the combination of an X-ray detection,  matched to IRAC sources which where not classified by \citet{Gut09} but had SEDs slopes consistent with Class II or shallower.  Most of these objects lacked either JHK or [5.8] and [8.0] data required by the multi-color algorithm.  We also required the [3.6] and [4.5] mags to be brighter than 16.  This latter requirement should exclude background AGN.  We combine the standard set with this augmentation which gives us 155 stars in the {\it augmented set}  (Table~\ref{TAB:STD}).  

In the overlap region, there are 21 variables among stars not in the standard set.  These variable stars are, in the mean, much fainter than the variables stars among standard set members (see Figure~\ref{depth_hist}). Standard set member variables have a median [3.6] magnitude $\approx$ 12 compared to about 13.5 for variable stars not included in the standard set.  
The majority of the SEDs of these variables, not in the standard set, indicate normal photospheres. 
This includes 7 objects identified by \citet{Gut09} as having normal photospheres and not detected in X-rays. 
Stars observed only in [3.6] generally, though not exclusively, lie to the north of the overlap field.  There are less than 20 stars in this field brighter than 13 and only 4 variables overall. 
Data from the southern region is primarily mapped in [4.5] where the median magnitude of the variables is  [4.5]=14.6. The SEDs of the 7 [4.5]-only  variables include 4 highly reddened objects  (2 of which would map as Class I if there is no line of sight reddening). 

From this we conclude that the northern and southern fields are very different from the overlap field.  In both cases, the variables are relatively faint and in the southern field they are more reddened. The southern field contains more ``legitimate'' cluster members as derived from the cryo maps \citep{Gut09}.  For both the northern and southern fields, we obtain a variability rate of $<$2.5\%. The rate of nearly 10\% seen among the non-standard set stars in the overlap region, combined with their collective brighter magnitude, indicates that most of the variables in the overlap region that are not part of the standard or augmented sets of cluster members by the procedure outlined in \S~\ref{sec:DR_SST} are, nonetheless, cluster members. Even including these sources does not make for a complete census in that we are only considering the $S >0.9$ and high $\chi^2$ objects as  variables.  We have already argued in \S~\ref{sec:noper} that  $S >0.9$ misses an unknown fraction weakly variable YSOs without a disk.

\section{Discussion}
\label{sec:Dis}

\subsection{Variability and Class}
The fraction of Class I and Class II objects found to be variable is nearly identical.  Class I/F sources show a higher variability rate, but Class II variables exhibit slightly larger changes  based on the median $\Delta$ [3.6] \& [4.5] values.  This is consistent with the results from \citet{Gun14} who found slight differences between rising, flat and Class II SEDs in L~1688.  There all 3 classes showed variability rates above 80\%, but a trend that the more positive the slope the higher the variability rate.  \citet{Gun14} also found the stars with the more positive sloped SEDs to have more intrinsic variability. 

\begin{figure*}[htbp]
\begin{center}
\includegraphics[height=7.5in]{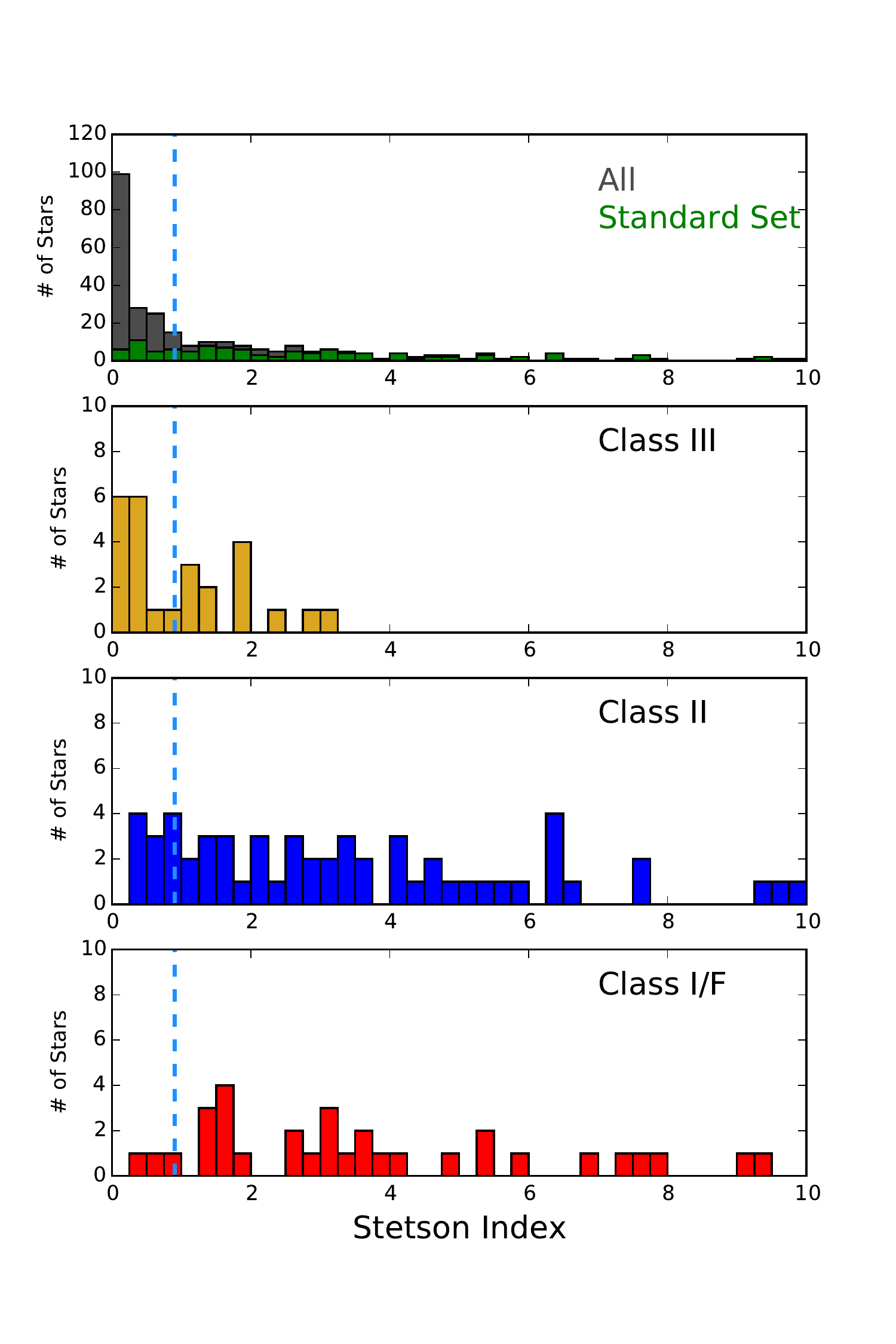}
\caption{ 
Histograms of Stetson indices.  The top plot shows a histogram of $S$ for all measured objects (grey) and the standard set (green).
The lower plots further break down the standard set by Class.   The vertical dashed line indicates our criteria for variability. At least a few Class I/II objects have $S<$0.5. 
}
\label{StetsonByClass}
\end{center}
\end{figure*}

The observed variability depends on Class. 
Class III stars  that vary  have observed changes of $\sim$ 7\% and are consistent with spotted stars as seen in other studies \citep{Car01, Wol13, Sch05}. 
At least 85\% of the Class I and Class II stars are variable. Our assessment is that neither  Stetson $> 0.9$ nor $\chi^2 > 5$ identify the weakest of the variables.  On the other hand, it is clear that the data are not consistent with 100\% of the Class I and Class II sources being variable over the roughly 40 day observation window since about 10\% of each group has a Stetson index below 0.5 (Fig~\ref{StetsonByClass}). Further, we find that magnitude is not a significant contributor to the relative variance observed.  We performed two-sided Anderson-Darling tests among the various subsets, for example the bright and faint stars from each of Class II and I.  That is, we compared The Stetson indices, $\chi^2$ and $\Delta$ values for Class II stars brighter than 13$^{th}$ magnitude with Class II stars fainter than 14.5 and find that the stars are consistent with having the same parent distributions. The Class III objects, on the other hand, are less variable, and found to be clearly different from the stars in the other classes.  

There are only three identified transition disk objects in the field as identified by \citet{Gut09}: SSTYSV J061035.92-061249.8,
SSTYSV J061036.97-061158.6 and
SSTYSV J061042.56-061513.9.
All three have SED slopes which identify them as Class II because, although they have weak excesses at $\lambda < 8$ \micron, they have strong excesses at 24\micron.  None of these were detected in X-rays.  The last of these three was only partially monitored in the [4.5] channel with only 29 measurements and was not in the [3.6] field. It was not noted as a variable.  The other two were both identified as periodic and hence both cataloged as variables.  But only SSTYSV J061036.97-061158.6 was identified as a variable via its $S$ index and $\chi^2$ values. As transition disks, these stars are expected to lack the inner disk material which is the hallmark of Class II objects and should vary in a manner more similar to diskless Class III objects at [3.6] and [4.5].  This is indeed true for SSTYSV J061036.97-061158.6 which has $\Delta$ of 3\% and 5\% for [3.6] and [4.5] respectively.  On the other hand, SSTYSV J061035.92-061249.8  has $\Delta$ of about 20\% in both channels. This may be an  AA~Tau type star in which the disk, periodically occults the star \citep{Bou03}. A number of the AA Tau variables in NGC~2264 have IR colors with transition-disk SED's \citep{Sta14,Sta15}.   This may be a projection effect as the IR SED is influenced by the view angle if it is close to edge-on.  These objects demonstrate that stars with transition-like SEDs do not necessarily vary like Class III's but
instead can have large light curve amplitudes due to variable extinction.  The fact that all three were not detected in X-rays 
and only one was detected as a Stetson variable indicates a significant undercount of diskless sources is likely even when including all variables in the overlap field. 

 In addition to changes in the individual channels, several groups  have interpreted the color change observed in the near-IR as a clue to   the physical nature of the change \citep[\ie disk accretion, extinction, spots, etc.;][]{Car01,Alv08, Wol13}. For all sources we calculated trajectories in the mid-IR color-magnitude plane. 
  We calculate slopes in the [3.6],[3.6]$-$[4.5] CMD such that 0$^{{\rm o}}$ = a color change toward the blue and no change in [3.6], 90$^{\rm o}$ = no color change and a decreased [3.6] flux. Trajectories peak around 85$^{{\rm o}}$.

\begin{figure*}[htbp]
\begin{center}
\includegraphics[width=6.5in]{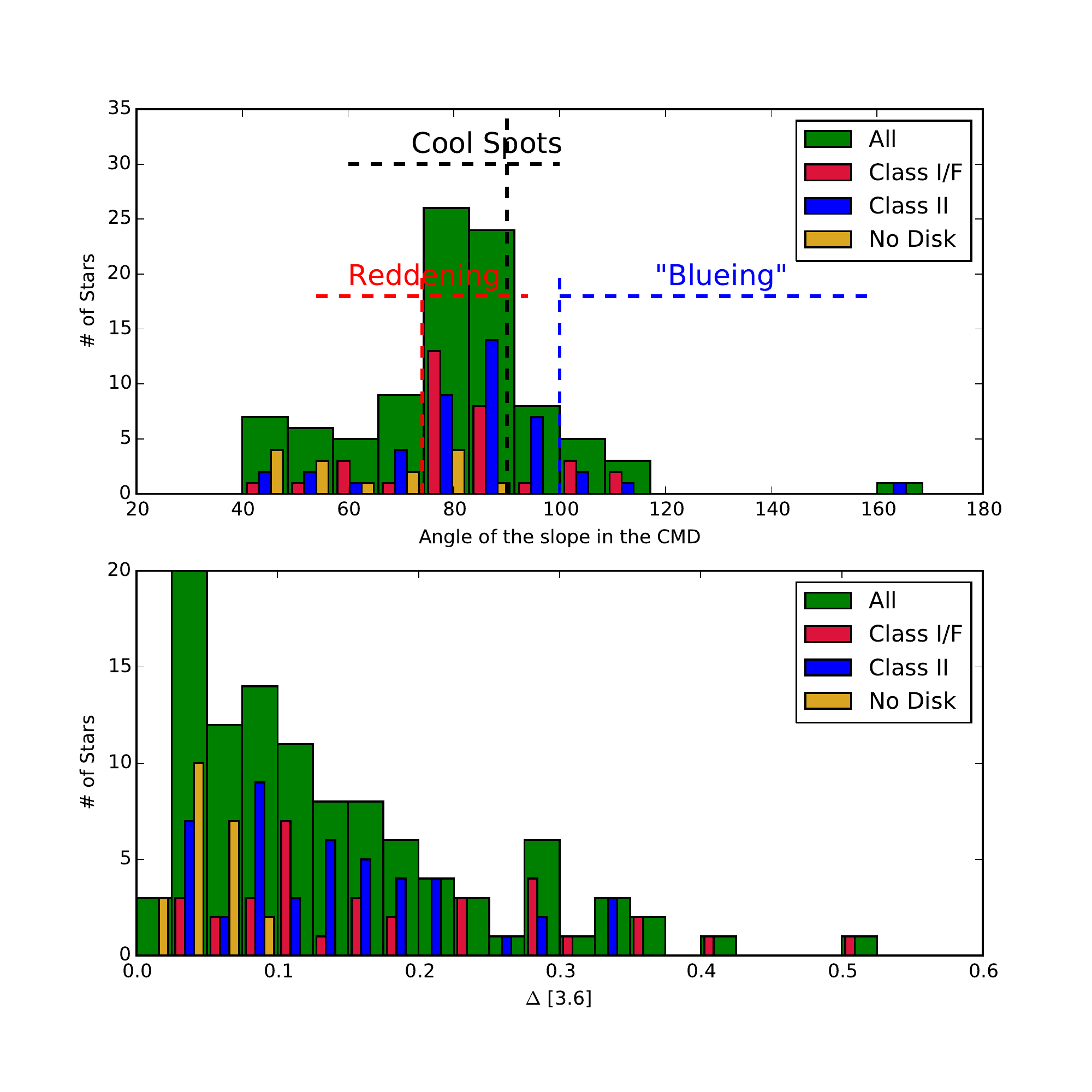}
\caption{ 
The fitted slope (top) and $\Delta [3.6]$ (bottom) for all the color-magnitude diagrams.  In this depiction, an angle of about 74$^{\rm{o}}$ corresponds to the reddening vector \citep{Ind05}. Cool spots tend to be colorless as they are mostly a geometric dimming at these wavelengths but can be red if they are very cool. Extreme accretion changes, including hot spots, can be blue since they are very hot. This effect is also seen in models presented by \citet{Car01} and observed recently in the near-IR \citep{Wol13}. Diskless objects show the smallest $\Delta [3.6]$, while Class~I and flat spectrum sources show the most extreme changes.}
\label{cmd_slopes}
\end{center}
\end{figure*}

There are about 100 variable sources brighter than median [3.6]=16 with a well-defined slope in color-magnitude space (\ie errors $<$ 6$^{\rm o}$). These are shown in Figure~\ref{cmd_slopes}.  The distribution of slopes is fairly symmetric with a full width at half maximum of about 15 degrees, which is consistent with the 3$\sigma$ slope measurement errors.  This main body of slopes is consistent with the expectation from reddening. Recasting the reddening derived by \citet{Ind05} as a slope in the [3.6], [3.6]-[4.5] plane, we derive a reddening slope of about 74$^{\rm{o}}$. However, the slope should be steeper if the dust is processed.  Similarly, cool spots, which have a change in luminosity with essentially no MIR color change, would have a slope very close to 90$^{\rm{o}}$. These can be hard to distinguish given an error of up to 6$^{\rm{o}}$. On the other hand, there is no physical mechanism for the slope  to exceed 90$^{\rm{o}}$ if it is due to either reddening or cool spots.\footnote{Fit errors could account for slopes angles as high as 100$^{\rm{o}}$.}   In fact, there is no reason for a YSO not to be impacted by {\em both} variable reddening and changing cool spot coverage.  

Slopes in the CMD, greater than 90 degrees indicate that the star gets bluer when it gets fainter \citep[See also][]{Fla13}. In analogy with the term reddening we describe this as ``blueing" \citep[][``bluening"]{Wol13,Gun14}.  There are 11 stars (about 10 \%) with slopes in excess of 100$^{\rm{o}}$ and hence indicative of blueing.  All of the blueing objects are disk bearing YSOs,  6 Class II objects, 4 flat spectrum objects and a Class I object.  Of these 11, six of the YSOs have fitted CMD slopes in the narrow range between 100-110$^{\rm{o}}$.  While 3$\sigma$ errors admit the possibility that these are due to dark spots, we believe that the measured slope is due to other factors as all have $\Delta$[3.6] $>0.08$.  This level of variability is uncommon in cool spotted stars, and the fits to the slopes in the CMD are very robust -- only one has errors $>3.5^{\rm{o}}$.   

Blueing cannot be accounted for by cool spots or disk eclipses, thus the phenomena has been suggested to relate to changes in the disk \citep{Wol13, Car01}.  One optical example of this effect  is the Herbig Ae star UX Ori which is seen to get bluer as the star fades from 
outburst.  This is explained by an increased fraction of scattered light contributing to the observed flux in the faint state.  For this mechanism to work in the IR, it would require a large fraction of the IR radiation to be scattered light and not intrinsic emission \citep{Gun14}. Almost half the blueing sources identified have flat or Class I SEDs and hence the IR emission is expected to be intrinsic. 

  It is also unlikely the blueing in the mid-IR is directly tied to accretion.  \citet{Fae12} find no relation between mid-IR light curves and spectroscopic accretion tracers.  \citet{Fla12} interpreted these kinds of changes as evidence of scale height variations. In the case of LRLL~21, they combined [3.6] and [4.5] photometry with IRS spectra and found that although the disk gets redder as the flux increases, the dominant change in the system is in the strength of the disk emission, not the temperature of the dust.
They assumed the stellar flux is fairly stable (i.e., the star might be spotted but has no flares) on the timescale which the infrared flux varies and that the radius of the inner disk should be constant. In this case, the changing flux implies that the emitting area of the disk is varying, and if the radius of the inner disk is constant then the scale height must be varying. \citet{Gun14} point out that when the inner disk wall puffs up, this will cast a shadow reducing the total emission at longer wavelengths. 
Alternatively,  \citet{Kes13} demonstrate that warped disks can reproduce features of the observed light curves.
Finally, \citet{Fla13} pointed out that the blueing effect could arise from any significant brightening of the stellar photosphere. In this case, there would be a larger stellar flux in [3.6] versus [4.5] diluting some of the change in [3.6] due to the change in disk flux when compared to the change in [4.5].   All the models are undoubtably simplistic and it is probable that individual cases have different root causes.


 The top portion of Figure~\ref{cmd_slopes}  indicates most stars show either neutral color changes or reddening, but a significant minority, mostly Class II, get bluer when they get fainter.   This $\sim$ 10\% minority is less than the number of blues seen by to \citet[][their Fig.\ 16]{Gun14} who find about half of all variables ``bluen'' in their  L~1688 study. 
 The lower plot shows $\Delta$ [3.6] for the same sources as the upper plot. Diskless objects indicate changes of less than 10\% - consistent with cool spots. Observed changes in the position on the CMD for Class II and Class I sources peak around 0.1 but there  is a long tail out to 0.5, and Class I sources show changes beyond 0.5 mag. 

\begin{figure*}[htbp]
\begin{center}
\includegraphics[width=6.5in]{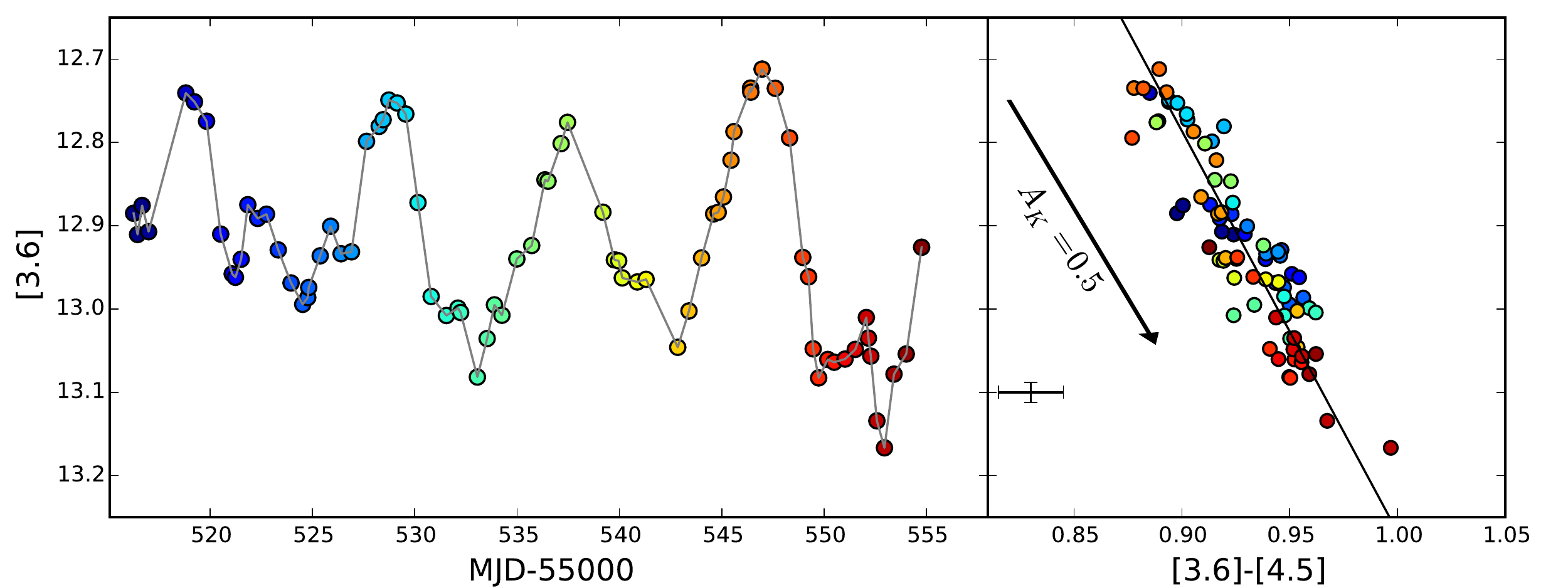}
\caption{Lightcurves for Class~II  object SSTYSV J061056.75-061101.7. In the left panel we show the raw lightcurve for the  [3.6] data.  In the right panel we show the color-magnitude diagram. The colors of the dots change with time as in the earlier figures. The fitted line is about 78$\pm1^{\rm {o}}$.
The $A_K$=0.5 reddening vector is indicated as are the measured slope and typical errors.  Notice the fitted slope and the reddening vector are almost parallel.
}
\label{fig:57186}
\end{center}
\end{figure*}

In Figure~\ref{fig:57186}, we show one of the most well behaved Class II reddening sources -- SSTYSV J061056.75-061101.7.  This star was found to have a 9.6 day period.  The changes in magnitude move in lock step with the color changes and are a very good fit to the ISM reddening vector \citep{Ind05}.  A variable, in which a Class~II star appears to be regularly occulted by disk material and dip in brightness is called an AA~Tau type, after the first such star modeled \citep{Bou03}.  How an AA~Tau-like appears in the mid--IR  is still somewhat ambiguous. In NGC 2264, only 10\% of the optical dippers, are also found to act similarly in the mid--IR.  In GGD~12-15 we find 4/39 objects in which the periodic change in color space has low errors ($<3^{\rm o}$) and is within 4 degrees of the ISM reddening vector in color space.  SSTYSV J061056.75-061101.7 is by far the cleanest. This is consistent with recent results from NGC~2264. There, $<$15\% of the variable stars showed quasi-periodic behavior in the IR which could be interpreted as AA~Tau-like \citep{Cod14}. 

\begin{figure*}[htbp]
\begin{center}
\includegraphics[width=6.5in]{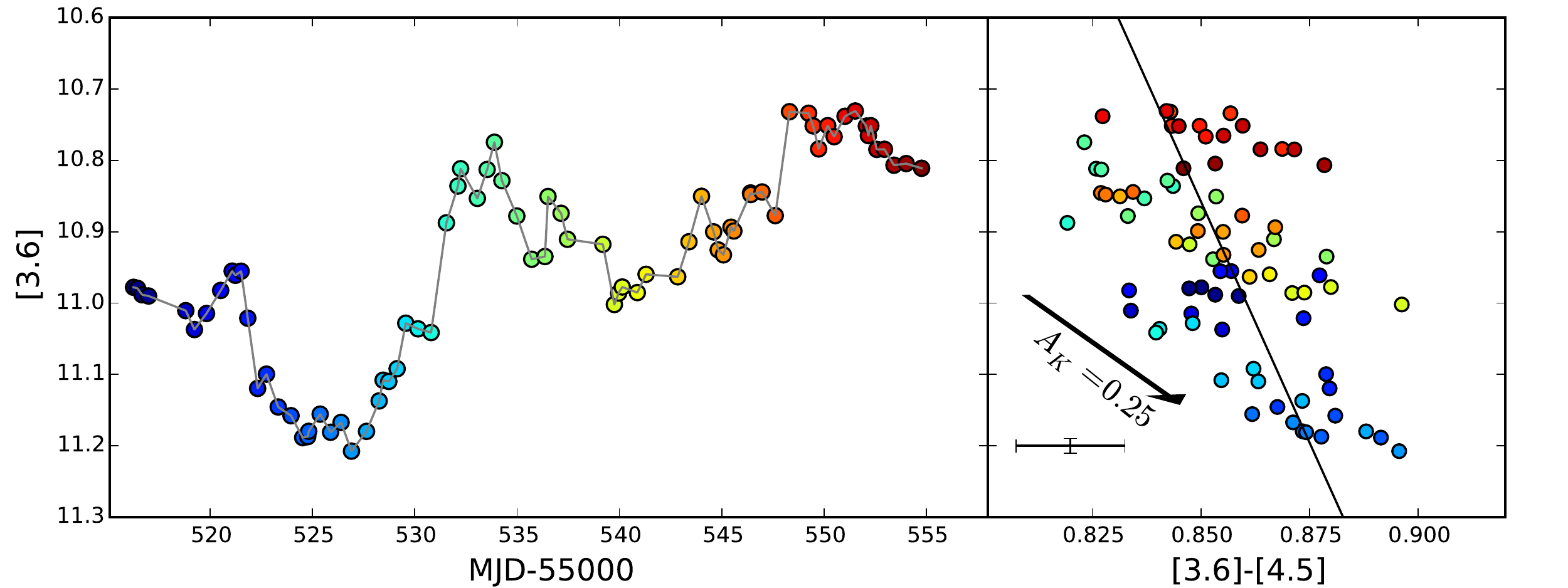}
\caption{ Same as above except for Class~II source SSTYSV J061047.70-061155.8. While the fitted slope is steeper than the reddening vector (in this case indicating  $A_K$=0.25), it is apparent  in the short term the trend in the data follows the reddening vector in the CMD. But over time the whole system brightens by about 20\% and the fitted angle to the color CMD is about 86 $\pm1^{\rm {o}}$.
}
\label{fig:57099}
\end{center}
\end{figure*}

 Figure~\ref{fig:57099}  shows  one of the largest color changes among Class II sources.  Again the slope is  dominated by reddening but here we note the large dispersion.    The dispersion is not random. SSTYSV J061047.70-061155.8  has a $K-[8.0]= 5$ and is one of the few stars detected at J through [8.0] bands. It starts relatively faint, gets fainter and then brightens steadily while remaining in a coherent color range. Another similar star is  SSTYSV J061053.41-061142.1, which is also an X-ray source with $K-[8.0] \sim 5$. It starts relatively bright and gets fainter with each consecutive period.
 Both of these objects seem to indicate a longer-term brightness change in addition to the observed week--scale variations.   

\begin{figure*}[htbp]
\begin{center}
\includegraphics[width=6.5in]{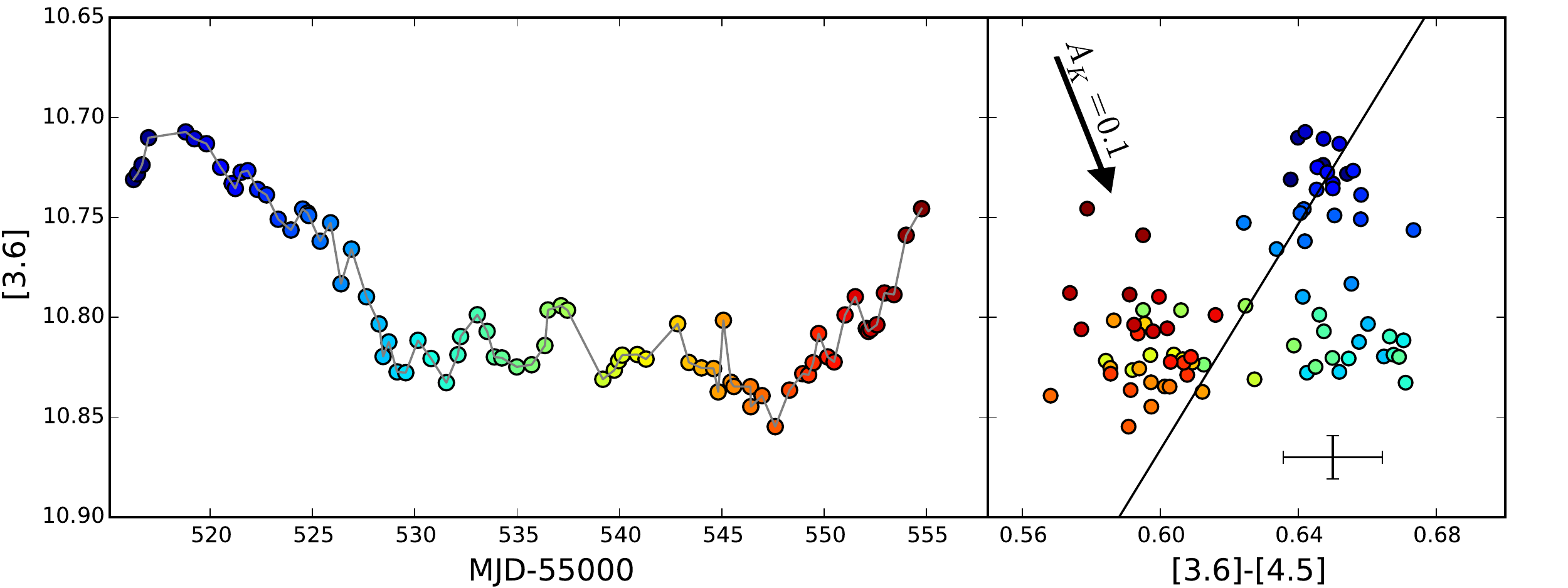}
\includegraphics[width=6.5in]{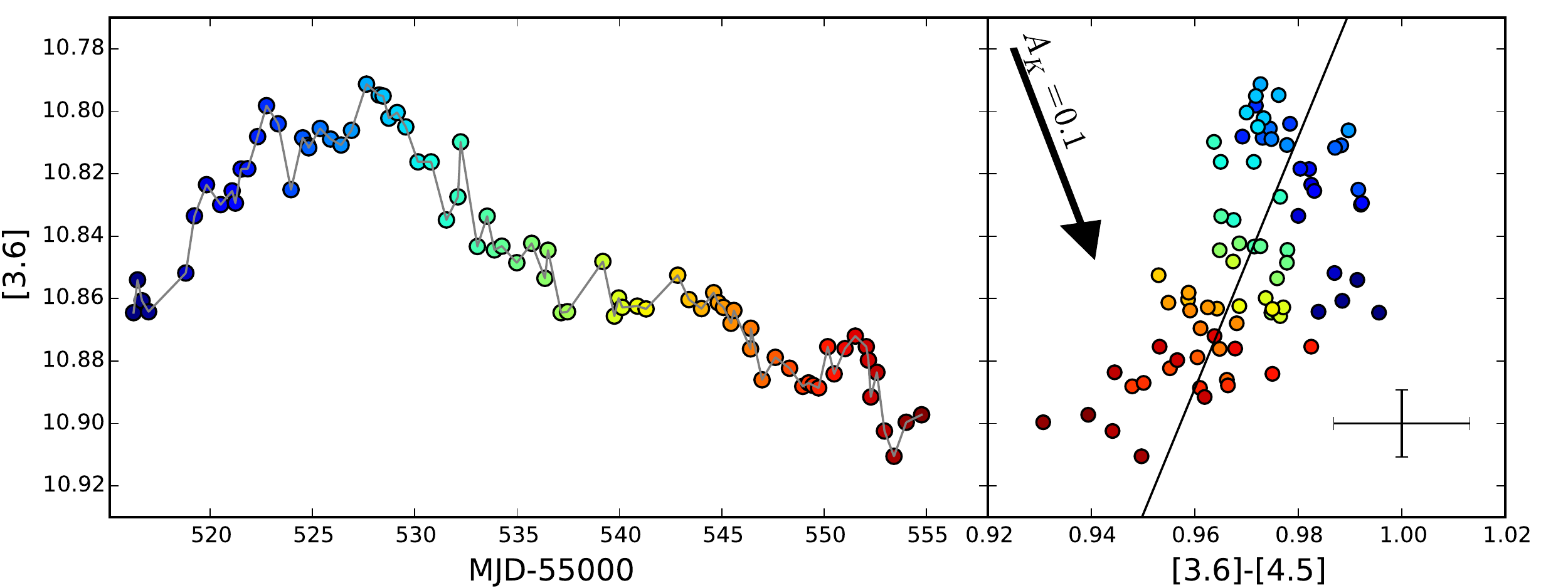}
\caption{ Class~II sources SSTYSV J061046.56-061304.4 (top) and SSTYSV J061048.76-061132.5 (bottom). These are two sources for which the fitted slope indicated significant blueing associated with the star becoming fainter.  For SSTYSV J061046.56-061304.4,  the various weeks of data cluster together indicating sudden changes. For SSTYSV J061048.76-061132.5,  The change is fairly steady  with the star traversing the same color space as it gets fainter as it does when it gets brighter trend. The $A_K$=0.1 reddening vectors and error bars are indicated.  The fitted slopes are  110$\pm4^{\rm {o}}$ and 104$\pm2^{\rm {o}}$  respectively. 
}
\label{fig:57090}
\end{center}
\end{figure*}

Figure~\ref{fig:57090}  shows the best two examples of blueing among the standard set.  SSTYSV J061046.56-061304.4 is a  Class II object with K-[8.0]~$\sim 3.5$.  There are three data clusters in the color--magnitude diagram.  The star is bright and red in the beginning, then shifts to 10\% fainter and still red for about a week and then shifts again, this time to the blue for the final 3 weeks.  The flat spectrum SSTYSV J061048.76-061132.5 has K-[8.0]~$> 4$. After the first week, its progression towards the fainter and bluer part of the color-magnitude diagram  is very steady, nearly monotonic.  It is not clear how long the trend continued.  

\subsection{Impact of X-ray on IR variability}
\label{sec:DisX}
Following the results obtained among the variable sources, we examined whether the presence of an X-ray detection affected the metrics we expect could be related to the type of object being monitored.  These impacts could include flares in the X-ray giving rise to flux increases or color changes in the IR.  We find, for example, that the observed X-ray variability (as measured by GL-vary) was similar for the Class II and Class III sources. Overall, there do not appear to be any correlations between the X-ray variability observed during the 20 hour X-ray observation and the Stetson index as measured across the 38 days of monitoring. Class II and Class III objects both range from no to moderate X-ray variability.   The three Class I sources for which GL-vary was measured all show very low X-ray variability, but the sample size is too small to draw any significant conclusions.  

Following on the result that disked sources have slower rotation periods  than non-disk sources, and that this is thought to be accounted for by coupling of the star to the disk via the magnetic field,  we wanted to test whether this kind of magnetic interaction persisted in the variable sources as a whole.  Using an Anderson--Darling test, we found that the distribution of  Stetson indices for X-ray and non-X-ray detected sources differed at about 97\% confidence.   However, as noted earlier, the Stetson index is a relative measure and hence very sensitive to the number of samples and inherent errors. So the brightness effects are critical. Indeed, we find that the X-ray sample is much brighter in the mid-IR than the typical sample of Class II's.  When this effect is accounted for by limiting the control group to Class II objects without X-rays that are brighter than [3.6]=12,  the significance of the Stetson effect vanishes. 

\begin{figure*}[htbp]
\begin{center}
\includegraphics[width=6.5in]{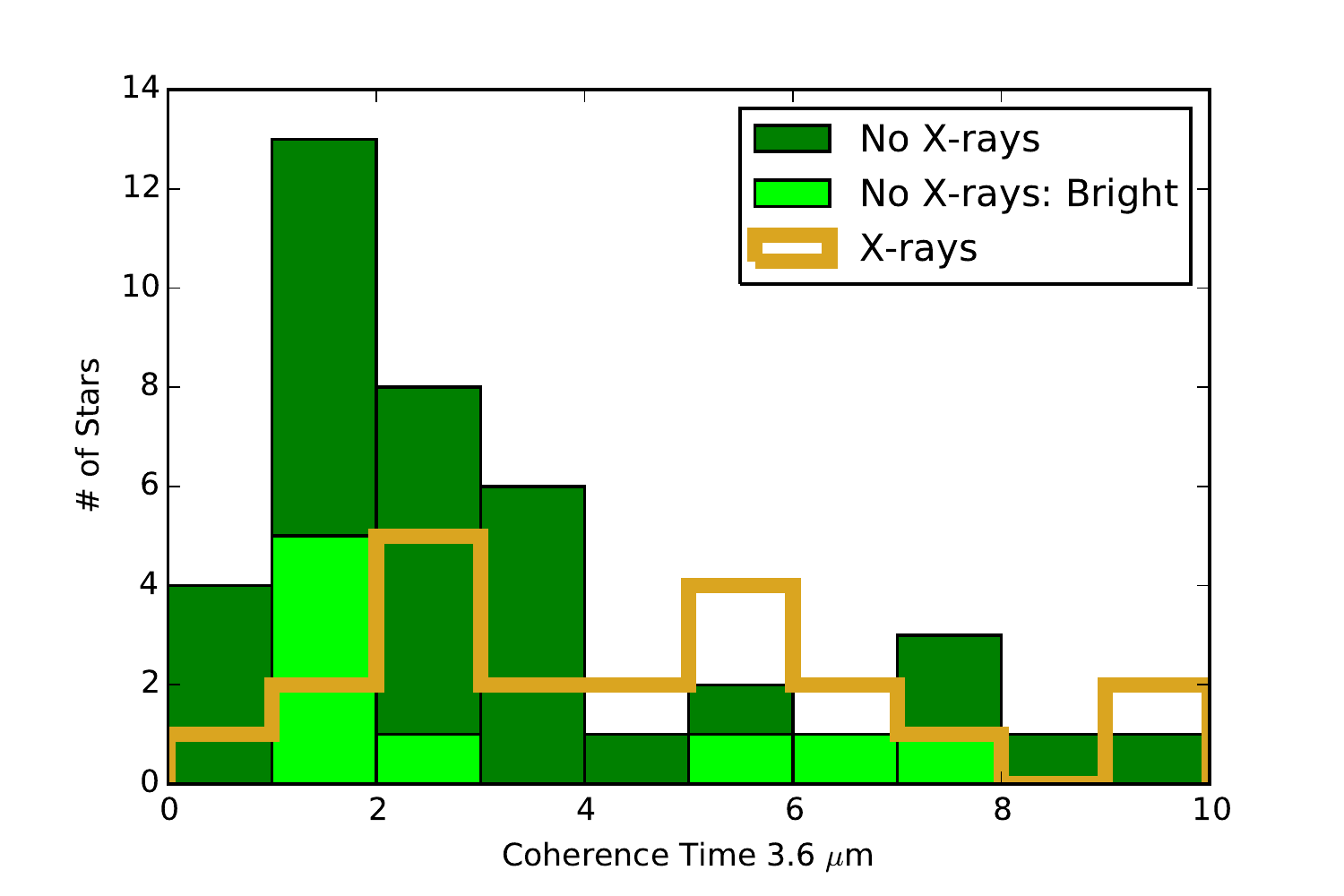}
\caption{ The distribution of coherence time for Class II sources, divided into non-X-ray and X-ray sources. The non-X-ray sources are further subdivided into a bright group brighter then [3.6]=12.  Coherence times tend to be longer for the X-ray sources. 
}
\label{fig:Coher}
\end{center}
\end{figure*}

On the other hand, the distribution of coherence time showed a significant trend.  In this case the Anderson--Darling test
showed that coherence time scales are different at 99\% confidence levels for the X-ray and non-X-ray selected samples. 
The X-ray sources show longer overall coherence time scales by a factor of about 2 (in the mean; Fig~\ref{fig:Coher}). 
This time, when the brightness selection effect is accounted for by limiting the control group to Class II objects without X-rays that are brighter than [3.6]=12,  the significance of the coherence effect is enhanced.

As described earlier,  the coherence time is a decay time of the value of the autocorrelation function.  Coherence time is a measure of how long is takes a star to significantly change. It is not the duration of a full cycle,  but is instead a measure of coherence that appears linearly correlated with the time period (see Fig.~\ref{periodVACF} and \citet{Pop15}).  There is wide scatter and the details of this are beyond the scope of this paper.  The point is that the coherence time can be related to the time scale of a non-periodic source and allows us to compare the time scales of non-periodic sources.
We could only compare Class II objects as they were the only group with a significant number of X-ray and non-X-ray detected objects (57 and 30 respectively). With the exception of coherence time, we find that in most obvious ways, i.e., magnitude change and color change ($\Delta$ 3.6, $\Delta$ 4.5, $\Delta$ [3.6-4.5]), X-ray and non-X-ray sources are consistent with being drawn from the same parent distribution.

\skipthis{One of the original goals of this observation was to determine if X-ray variability (\eg\ flares) did or could induce mid--IR variability.  Alas, we could not achieve this goal for at least two reasons.  
As in \citet{Fla14}, none of the X-ray sources had a strong flare, and there were only 4 weak flares. Conversely, the scale of the observed variability in the mid-IR was fairly high, typically 10\%.  
\citet{Ke12} model YSOs fairly extensively to test the effect of X-rays and especially X-ray flares on IR variability.  While they find a fairly short response time for a brightening below 5\micron, the model uses a flare of 10$^{32}$erg sec\minusone\ to get a significant effect.  This is a factor of 100 brighter than anything observed in this observation of GGD~12-15. }

\subsection{GGD 12-15 in the YSOVAR Context}

GGD 12-15 is one of 11 clusters included in the project called YSOVAR which is outlined in Paper~I.  There is also a related projected focused solely on NGC 2264  which employs simultaneous optical and IR data \citep{Cod14,Sta14, Sta15}. 
The current goal of the YSOVAR project is to perform and present a detailed analysis of each cluster. Once each individual cluster has been understood in its own context, there will be a focus on the ensemble properties. This is the purpose of that standard set of cluster members. So it is already worthwhile to see how GGD 12-15 compares with the other clusters for which detailed studies have been completed.  

Completed mid-IR variability studies exist for four other clusters: IC 1396A \citep{Mor09}, NGC 2264 \citep{Cod14,Sta14, Sta15}, Lynds 1688 \citep{Gun14} and IRAS 20050+2720 \citep{Pop15}. There is also a preliminary overview of all clusters (Paper~I)  and a cursory report on a few sources in Orion \citep{Mor11}.  All analyses find very high levels of variability for YSOs -- especially those with IR excesses relative to a plain photosphere.   In this section, we will focus on comparing the results from the first four clusters which have been analyzed in a manner similar to GGD~12-15. 

L~1688 was the subject of the first of the full YSOVAR cluster studies. This cluster is highly embedded and ranks as the second youngest of the YSOVAR clusters in terms of class ratios (the ratio of  Class I to all YSOs or of Class I to Class II; See Figure 6 of Paper~1). The advantage of this cluster was the fairly complete cluster census due to its close distance ($\sim$ 140 pc) as well as a long study baseline covering several seasons.  However this also meant that the fields are relatively sparse and relatively few cluster members are studied.  Among the known cluster members, about 90\% were identified as variable.  For cluster members, there is a clear correlation between evolutionary status and IR variability. More embedded sources are more often detected to be variable, and they have larger variability amplitudes.   These were measured as $\Delta$ [4.5] =0.14 mag, 0.13 mag, 0.12 mag, 0.08 mag for Class I, flat spectrum, Class II and Class III respectively.  About 20\% of the cluster members were found to be periodic variables.  It was also noted that at least 6 variables get bluer when they get fainter, not redder -- as would be expected by dust obscuration. The blueing sources were never Class~III.  One source, ISO-Oph 140, traversed a loop in color-magnitude space -- we saw no corollary for that behavior in GGD~12-15. No correlations were seen between X-ray and variability characteristics in L~1688  but there were no contemporaneous X-ray observations. 

There other clusters investigated so far have been more distant. IRAS 20050+2720 and NGC 2264 are of similar age in terms of class ratio to L 1688.  But their distance  (about 700 parsec) is closer to that of  GGD 12-15.  In IRAS 20050+2720 \citep{Pop15}, a little more than 70\% of all YSOs were found to be variable.  The biggest difference between IRAS 20050+2720  and the other clusters is among Class I sources where only 65\% were found to be variable. Only some of this is due to luminosity issues.  Even when the sample was limited to bright sources, only 71\% of the Class I sources were found to be variable.  On the other hand, about 85\%  of the flat spectrum and Class II sources in IRAS 20050+2720 are found to be variable.  The sense of this is the opposite of what is seen in GGD 12-15 wherein the steeper SEDs are more likely to be found to be variable. The binomial errors on both distributions are about 5\% so the difference appears significant in both cases. 

While the lower variability rate in IRAS 20050+2720 appears robust, it  does not appear to be common.  \citet{Cod14} report 91\% of 162 disk bearing YSOs in NGC 2264 are variable from the mid-IR perspective.  The $Spitzer$ observations of NGC 2264 were very dense with about 300 observations over the course of a month.  This allowed the use of a very tight variability threshold of $S> 0.21$ and correspondingly, more weak variables may have been identified.  But this is probably a small effect as many of the weaker $S$ variables have other variability indicators including periodicity.  

This latter point can be demonstrated by a detailed examination of IC 1396A.  Data for this cluster were taken over a 14 day span in all 4 IRAC channels.    \citet{Mor09} used a $\chi^2$  p value of 0.01 (equivalent to about $\chi^2$= 6.7 per degree of freedom) which is a little greater than the value we have settled on.  They also used an intermediate value of $S$=0.55 as the variability threshold.  This difference means it is possible some of the weaker variables identified by \citet{Mor09} in IC 1396A  would not have been identified as variables by the methods used here. Fortunately, inspection of Table~2 in \citet{Mor09} found no sources which would have been classified differently with regard to their variability despite the minor differences in the technique used. 
Variability tests based on the [3.6] and [4.5] data show a variability fraction of just under 65\% for stars with strong IR excesses. We attribute this to the shorter baseline.
 \citet{Ric15} show that variability tests, in the near-IR are sensitive to the time baseline used. They recover $\sim$ 5-10 \% more variables in 40 day samples compared to 14 day samples on the same JHK dataset.  This is both an effect of improved signal to noise and that median K amplitude of the stars themselves increased as the observation window increases. Both \citet{Ric15} and \citet{Pop15} find that the timescale in bluers is longer than the time scale of reddening sources. So the lower variability rate seen in IC 1396A is, at least in part attributable to the shorter time baseline, but more work needs to be done to quantify this effect.  The duration of the IC 1396A study was too short for solid period determinations but 18/69 YSO are found to have period-like changes.  That is a higher rate of periodicity than seen in L1688 but very close to what we find  in GGD~12-15 Overall, the key trend of the high variability rate among disked sources remains.

\section{Summary}
\label{sect:summary}

We present $Spitzer$ \& $Chandra$ observations of YSOs in the star forming region GGD~12-15. There are 79 IR observations taken over a 38 day period observed with a non-uniform cadence to prevent bias in period detection, plus a moderately deep 67 ks $Chandra$ X-ray observation. 
Our sample consists of 1017 sources with lightcurves in [3.6] or [4.5] with at least 5 data points. The IR data divide into 3 subsets;  342 sources observed in the [3.6] channel -- mostly to the north;  a subset of 319 sources observed in the [4.5] channel, and
  an overlap  field with 356 sources observed in both channels.  Of those sources, we classify 149 sources as variable either because they are periodic or determined to be variable via the Stetson test or a $\chi^2$ test. The vast majority of variables are in the overlap field.  By comparing the variability levels, we determine that the northern and southern fields are essentially background fields. 
There were also 90 X-ray sources identified with the IR sources.  The X-ray sources include about 30 previously known Class~II and 5 Flat spectrum or Class~I sources and 39 possible Class~III sources. We define as a standard set of members YSOs previously identified by \citet{Gut09} plus the X-ray sources which are coincident with normal photospheres.  The standard set of cluster members totals 141 sources including 106 variables.     The X-ray data allow us to identify 14 additional Class~I and Class~II candidates.   The remaining stars in the overlap field also show a high variability rate indicating up to 21 additional YSOs among the variables in the overlap field.

Key findings include:
\begin{itemize}
\item Over 85\% of all Class~I, flat spectrum, and Class II sources exhibit high levels of (usually) correlated variability at 3.6 and 4.5 \micron.  This is very close to the variability levels reported for disked objects in NGC~2264 and L~1688.   The variability rate seen in GGD~12-15 is higher than for the disked objects in IC 1396A, but that can be accounted for by the longer observing window available here. IRAS 20050+2720 also shows a significantly lower variability rate. 

\item Three out of 36  Class I and flat spectrum sources are non-variable over the period of the observations.  Ten percent non variability for the flat and rising spectrum objects seems typical. This is exactly the rate found for these subclasses in L1688 which was the most heavily monitored of the YSOVAR clusters studied in depth so far.  Flat spectrum and Class~I sources in IRAS 20050+2720 and IC 1396A have non-variability rates closer to 25\%. 

\item Class I sources show both the largest fraction of variables and the largest brightness changes.  For this subclass,  we find median changes of 0.15 and 0.14 ($\Delta 3.6$ and $\Delta 4.5$ respectively). Flat spectrum sources show the second highest variability fraction, but the scale of the changes observed is statistically indistinguishable for all 3 classes of disk bearing YSOs  ($\Delta 4.5$ = 0.13 for both Class~II and flat spectrum objects).  The results in L~1688 and IRAS 20050+2720 are almost identical and NGC 2264 shows a similar trend.

\item Just under half of the Class III sources are identified to be variable. This includes several periodic sources which would not otherwise have been noted as variable, as they had Stetson indices and $\chi^2$ below our thresholds for variable designation. This is the hardest grouping to compare among the clusters as most of them have different biases. In the only cluster which used the identical source identification procedure, IRAS 20050+2720,  the same result is found. 


\item The Class III sources show  less change in color than the other classes with little change in color.  The maximum color change of a Class~III is $\Delta$[3.6]-[4.5]$ \sim 0.08$. There are many Class~II -- Class~I sources which become redder when they dim suggesting (a)periodic disk obscuration. The maximum  $\Delta$[4.5] measured for a standard set variable was a Class~I source with $\Delta$[4.5] = 0.48.

\item Nine sources out of 101 variables with well measured slopes in the CMD get significantly bluer as they dim. All 9 are disk-bearing YSOs.  This is similar to the fraction for L 1688 and IRAS 20050+2720.  In near-IR colors (\eg, K vs. H$-$K) the fraction of bluers  approached one-third.  Following \citet{Wol13}, this is suggestive of an accretion event, but many scenarios are possible. In addition, several sources show compound trends; for example, dominated by reddening but with an orthogonal motion in the [3.6], [3.6]$-$[4.5] color-magnitude diagram.

\end{itemize}

At the outset, a goal for this project was to relate X-ray flares with MIR variability.   Unfortunately, there were no strong X-ray flares among the 90 X-ray/IR sources.  There were no obvious trends seen  regarding X-ray brightness, X-ray variability or even X-ray detection with the observed IR variability. On the other hand, there was a $\sim$ 3$\sigma$ signal that the coherence time for X-ray detected Class~II sources was longer than the coherence time for non-X-ray detected Class~II.  While not definitive, this is the first direct statistical result identifying an infrared property as being dependent on the X-ray flux.

\acknowledgements
This work is based on observations made with the \textit{Spitzer} Space Telescope, which is operated by the Jet Propulsion Laboratory, California Institute of Technology under a contract with NASA. Support for this work was provided by NASA through an award issued by JPL/Caltech.  This research made use of Astropy, a community-developed core Python package for Astronomy \citep{Ast13}. This research has made use of the SIMBAD database and the VizieR catalogue access tool \citep{Och00}, both operated at CDS, Strasbourg, France and of data products from the Two Micron All Sky Survey, which is a joint project of the University of Massachusetts and the Infrared
Processing and Analysis Center/California Institute of Technology, funded by the National
Aeronautics and Space Administration and the National Science Foundation. H.M.G. acknowledges \textit{Spitzer} grant 1490851. K.P was funded through the Sagan Fellowship program executed by the NASA Exoplanet Science Institute. 
H.Y.A.M.\ and P.P.\ acknowledge support by the IPAC Visiting Graduate Fellowship program at Caltech/IPAC.
P.P.\ also acknowledges the JPL Research and Technology Development and Exoplanet Exploration programs.
RAG gratefully acknowledges funding support from NASA ADAP grants NNX11AD14G and NNX13AF08G and Caltech/JPL awards 1373081, 1424329, and 1440160 in support of \textit{Spitzer} Space Telescope observing programs. S.J.W. was supported by NASA contract NAS8-03060.  We thank Fabio Favata and David James for critical readings of this paper.

{\it Facilities:} \facility{Spitzer}, \facility{Chandra}

\end{document}